\documentclass[pdflatex,sn-mathphys-num,iicol]{sn-jnl}

\usepackage{graphicx} 
\usepackage{multirow}%
\usepackage{amsmath,amssymb,amsfonts}%
\usepackage{amsthm}%
\usepackage{mathrsfs}%
\usepackage[title]{appendix}%
\usepackage{xcolor}%
\usepackage{textcomp}%
\usepackage{manyfoot}%
\usepackage{booktabs}%
\usepackage{algorithm}%
\usepackage{algpseudocode}%
\usepackage{heppennames2}      
\usepackage{ECFAreport_definitions}
\usepackage{multicol}

\newcommand{\ab}{\,ab$^{-1}$}
\newcommand{\eqn}{equation}
\newcommand{\lam}{\lambda}
\newcommand{\lb}{\left (}
\newcommand{\rb}{\right )}
\newcommand{\sqs}[1]{$\roots=$#1\,\GeV}
\newcommand{\mch}{$M_{H^{\pm}}$}
\newcommand{\ma}{$M_A$}
\newcommand{\mh}{$M_H$}
\newcommand{\msmh}{$M_h$}
\newcommand{\mdiff}{\ensuremath{M_A-M_H}}
\newcommand{\mytt}{$\mathrm{t}\bar{\mathrm{t}}$}
\newcommand{\myqq}{$\mathrm{q}\bar{\mathrm{q}}$}
\newcommand{\ET}{\ensuremath{E_\text{T}}\xspace}
\newcommand{\etmiss}{\ensuremath{E_\text{T}^{miss}}\xspace}
\newcommand{\emiss}{\ensuremath{E^{miss}}\xspace}
\begin{document}

\title{Search for additional scalar bosons within the Inert Doublet Model in a final state with two leptons at the FCC-ee}


\author[1]{\fnm{Anubha} \sur{Bal}}
\author[1]{\fnm{Edward} \sur{Curtis}}
\author*[1]{\fnm{Anne-Marie} \sur{Magnan}}\email{a.magnan@imperial.ac.uk}
\author[1]{\fnm{Benedikt} \sur{Maier}}
\author*[2,3]{\fnm{Tania} \sur{Robens}}\email{trobens@irb.hr}
\author[1]{\fnm{Nicholas} \sur{Wardle}}


\affil*[1]{\orgdiv{Physics Dept.}, \orgname{Imperial College London}, \orgaddress{\street{Prince Consort Road}, \city{London}, \postcode{SW7 2BW}, \country{United Kingdom}}}

\affil[2]{\orgname{Rudjer Boskovic Institute}, \orgaddress{\street{Bijenicka cesta 54}, \city{Zagreb}, \postcode{10000}, \country{Croatia}}}

\affil[3]{\orgdiv{Theoretical Physics Dept.}, \orgname{CERN}, \orgaddress{\city{Geneva 23}, \postcode{1211}, \country{Switzerland}}}

\abstract{\noindent


 We present a search for new scalar bosons predicted by the Inert Doublet Model
at an $e^+e^-$ machine with centre-of-mass energies $\roots$ of 240 and 365 \GeV. Within this model, four additional scalar bosons ($H,\, A,\, H^+$ and $H^-$) are predicted. Due to an additional symmetry, the lightest new scalar, here chosen to be $H$, is stable and provides an adequate dark matter candidate. The search for pair production of the new scalars is investigated in final states with two electrons or two muons, in the context of the future circular collider proposal, FCC-ee. Building on previous studies in the context of the CLIC proposal, this analysis extends the search to detector-level objects, using a parametric neural network to enhance the signal contributions over the Standard Model backgrounds, and 
 sets projected exclusion and discovery contours in the \mdiff\ vs. \mh\  plane.
With a total integrated luminosity of 10.8\,(2.7)\,\ab\ for \sqs{240\,(365)},  the discovery reach for the model goes up to \mh$= 108\,(157)\,\GeV$ for \mdiff$=15$\,\GeV\  at \sqs{240\,(365)}. For exclusion, 
almost the entire phase-space available in the \mdiff\ vs \mh\ plane is expected to be ruled out at 95\% CL, reaching up to \mh$=110\,(165)$\,\GeV. 

\noindent
RBI-ThPhys-2025-08, CERN-TH-2025-069

}

\maketitle

\graphicspath{ {./figs/} }


\section{Introduction}

Despite the numerous cosmological observations pointing towards the existence of dark matter (DM) \cite{ParticleDataGroup:2024cfk}, the specific nature of DM is still an ever elusive mystery. This lack of understanding is further compounded by the fact that the abundance of DM is predicted to significantly outweigh that of normal matter. Assuming DM can interact directly with normal matter, it should be possible to detect it here on Earth. This can happen in a number of ways: DM passing through our solar system could directly interact with particles, imparting energy that can be detected in direct detection experiments; DM particles could modify Standard Model (SM) interactions and be detected indirectly through precision measurements; or, DM can be produced at a collider, leading for example to a missing energy signature. More details can be found in Ref.~\cite{ParticleDataGroup:2024cfk}.

We consider the case in which DM particles are weakly interacting massive particles. A particle of this nature, however, is not contained in the SM, and as such alternatives theories --- or extensions to the SM --- are required to include this. In the context of extensions to the SM such as two-Higgs-doublet models, it is possible, via the addition of an exact $\mathbb{Z}_2$ symmetry, to make the lightest scalar boson from the additional doublet stable, and hence a suitable DM candidate. The study presented here considers a search for one such theory: the Inert Doublet Model (IDM) \cite{Deshpande:1977rw,Barbieri:2006dq,Cao:2007rm,Kalinowski:2020rmb,Ilnicka:2015jba}.

Although many phenomenological studies exist, dedicated studies of the IDM have not yet been performed by the experimental collaborations. Existing analyses are generally not sensitive to a large part of the parameter space due to too high requirements on the missing transverse energy, and their recast leads to very loose constraints (see e.g. discussion in \cite{Dercks:2018wch}). In this paper, we present a search for the pair production of the extra scalars in $e^+e^-$ collisions at centre-of-mass energies ($\roots$) of 240 and 365\,\GeV~ in the context of the future circular collider of CERN, FCC-ee~\cite{FCC:2018evy,Blondel:2021ema,Agapov:2022bhm,FCC:2025lpp}. Within the IDM, the main target process is the pair-production $e^+ e^-\,\rightarrow\, A H$, where $A$ and $H$ are additional new physics scalars and $H$ serves as the dark matter candidate. The scalar $A$ subsequently decays into $Z\,H$, and we use a final state with one same-flavour, opposite-sign lepton pair (electrons or muons) accompanied by missing energy from the invisible dark scalars.

Similar studies for this channel have already been performed in previous work within the CLIC environment, where higher centre-of-mass energies are also investigated, see e.g. \cite{Kalinowski:2018kdn,Zarnecki:2020swm}. In this work, we extend on the previous studies by using full reconstruction. We also extend the interpretation to include the full kinematically-available mass ranges in a systematic way, thanks to the use of a parametric neural network.

This paper is organised as follows: the theoretical model is introduced in section~\ref{sec:idm}. The signal and background Monte Carlo samples are detailed in section~\ref{sec:mc}, with the object definitions given in section~\ref{sec:setup}. Preselection and selection criteria that enhance the signal contribution and reject background processes are defined in section~\ref{sec:sel}, and a multivariate analysis to further discriminate between signal and backgrounds is introduced in section~\ref{sec:pnn}. Finally, the results are given in section~\ref{sec:res} and a conclusion in section~\ref{sec:concl}.

\section{The Inert Doublet Model}
\label{sec:idm}

The Inert Doublet Model (IDM) is a two-Higgs-doublet model with an additional, unbroken $\mathbb{Z}_2$ symmetry. This model contains two scalar fields denoted $\phi_D$ and $\phi_S$, with the following transformation properties:

\begin{equation}
    \mathbb{Z}_2 : \, \phi_D \longrightarrow -\phi_D, \;
    \phi_S \longrightarrow \phi_S, \; SM \longrightarrow SM.
\end{equation}

With these field of the IDM, the most general, renormalisable scalar potential can be constructed:

\begin{eqnarray}
        V &=& -\frac{1}{2}[m^2_{11}(\phi_S^\dagger\phi_S) + m^2_{22}(\phi_D^\dagger\phi_D)] + \frac{\lambda_1}{2}(\phi_S^\dagger\phi_S)^2 \nonumber\\&& + \frac{\lambda_2}{2}(\phi_D^\dagger\phi_D)^2 
        + \lambda_3(\phi_S^\dagger\phi_S)(\phi_D^\dagger\phi_D) \nonumber\\&&
        + \lambda_4(\phi_S^\dagger\phi_D)(\phi_D^\dagger\phi_S) + \frac{\lambda_5}{2}[(\phi_S^\dagger\phi_D)^2 + (\phi_D^\dagger\phi_S)^2]. \nonumber\\&&
\end{eqnarray}

Due to the additional symmetry, electroweak symmetry breaking proceeds as in the SM. After symmetry breaking, the model contains in total seven free parameters. A suitable choice for these is e.g.
\begin{\eqn*}
v,\,M_h,\,M_H,\,M_A,\,M_{H^\pm},\,\lam_2,\,\lam_{345}\,\equiv\,\lam_3+\lam_4+\lam_5,
\end{\eqn*}
where $v$ and $M_h$ are the vacuum expectation value (vev) and mass stemming from the SM-like doublet and are fixed through experimental measurements to $v\,\sim\,246\,\GeV$ and $M_h\,\sim\,125\,\GeV$, respectively. $H,\,A,$ and $H^\pm$ denote the novel so-called dark scalars, and $\lam_i$ are couplings in the potential with $\lambda_{345} \equiv \lambda_{3} + \lambda_{4} + \lambda_{5}$. Here,
$\lambda_{2}$ is the coupling constant for the quartic vertex between four IDM particles, and $\lambda_{345}$ is the coupling constant for the vertex between IDM particles and the SM Higgs boson. The couplings between the new scalars and the SM gauge bosons are determined solely by quantities from the SM electroweak sector. It should be noted that the scalars from the two doublets do not mix due to the $\mathbb{Z}_2$ symmetry, therefore there are no additional mixing parameters.

The $\mathbb{Z}_2$ symmetry leads to multiple interesting phenomenological features. First, the symmetry forbids any direct coupling between the IDM particles and the SM fermion fields, and as such the IDM particles only couple directly to bosons. Second, the symmetry requires that all terms in the Lagrangian involving IDM fields have an even number of IDM fields
(2 or 4). As the second doublet does not acquire a vacuum expectation value, the symmetry remains exact and IDM particles are always produced in pairs. Finally, it also causes the lightest of the IDM particles to be stable and therefore a suitable DM candidate\footnote{A detailed discussion on regions where the IDM can render the observed relic density can be found in e.g.~\cite{Kalinowski:2020rmb}.}.
In principle, any of the additional scalars can serve as a DM candidate. We here chose $H$, which automatically implies a mass hierarchy \ma,\mch$>$\mh.

\begin{figure}[h!]
    \centering
        \includegraphics[width=0.3\textwidth]{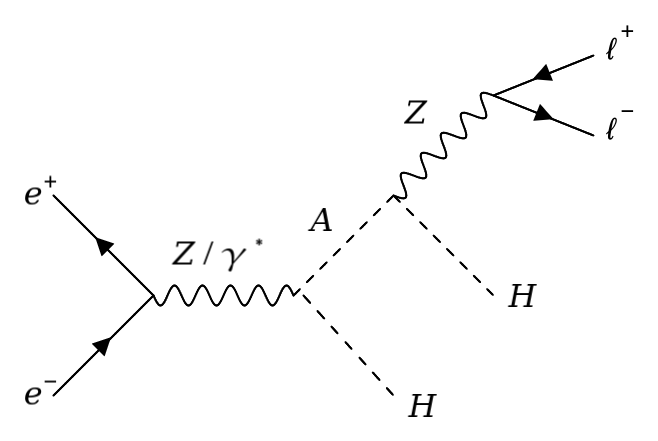}\hfill
        \includegraphics[width=0.3\textwidth]{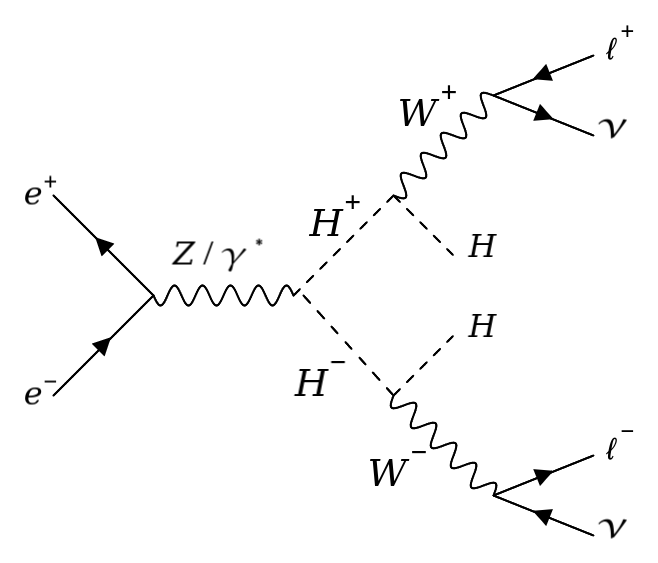}\hfill
        \includegraphics[width=0.3\textwidth]{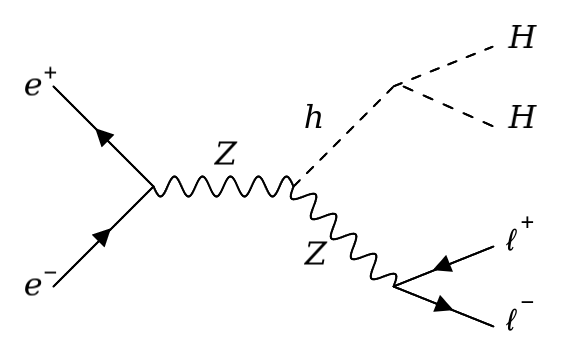}
    \caption{\label{fig:idm_diagram} Leading-order Feynman diagrams of: (top) $HH\ell^+\ell^-$ production,
        (middle) $HH\ell^+\ell^-\nu\bar{\nu}$ production, and (bottom) $Z h\rightarrow HH\ell^+\ell^-$ production, with $\ell=e,\mu,\tau$.
    }
\end{figure}

Given that the IDM particles do not couple to the SM fermions, a typical signature at a collider is that of SM electroweak gauge boson plus missing energy, where the latter stems from the DM candidates. In addition, the model can also render monojet and monophoton signatures~\cite{Belyaev:2018ext,Belyaev:2016lok} as well as processes containing the invisible decay of the 125 \GeV\ SM-like scalar. This analysis focusses on the same-flavour dilepton (electrons or muons) plus missing energy final state, with example production mechanisms shown in figure~\ref{fig:idm_diagram}. 

The IDM is subject to a large number of theoretical and experimental constraints. In this work, we follow the scan presented in Ref.~\cite{Ilnicka:2015jba}, with more recent updates presented in Refs~\cite{Ilnicka:2018def,Kalinowski:2018ylg,Dercks:2018wch,Kalinowski:2020rmb,Braathen:2024lyl}\footnote{Note we do not apply the two-loop constraints on the scalar couplings presented in \cite{Braathen:2024lyl}, but here chose to use the leading-order values for constraints.}. In particular, we include constraints on the potential stemming from vacuum stability, perturbativity and perturbative unitarity, as well as electroweak precision observables (EWPO), evaluated through the publicly available tool 2HDMC \cite{Eriksson:2009ws,Eriksson:2010zzb}. Results for the electroweak precision constraints are tested using the oblique parameters $S,\,T,\,U$ \cite{Altarelli:1990zd, Peskin:1990zt,Peskin:1991sw,Maksymyk:1993zm} and comparing to the latest PDG values \cite{ParticleDataGroup:2024cfk}. We also require that the parameter point resides in the inert vacuum \cite{Ginzburg:2010wa}. We furthermore test for the most recent bound on SM Higgs boson decays to invisible particles in cases where this decay is allowed onshell, $\text{BR}\lb h\,\rightarrow\,\text{inv}\rb\,\leq\,0.11$ \cite{ATLAS:2023tkt}. Regarding dark matter constraints, we calculate the predictions using Micromegas \cite{Belanger:2018ccd} and require agreement with the upper bound for relic density as measured by the Planck experiment \cite{Planck:2018vyg}. Direct detection constraints are taken from the LUX-ZEPLIN collaboration \cite{LZ:2024zvo}. Finally, we also make sure the mass hierarchy obeys the constraints from a recast of searches for supersymmetric neutralinos at LEP reinterpreted within the IDM \cite{Lundstrom:2008ai} (referred to as LEP SUSY recast in the following), and that the particles would not contribute to the decay widths of the electroweak gauge bosons at leading order. In the remainder of this document, we will label points as allowed that fulfill all above constraints, if not mentioned otherwise. More details on the scan setup and the applied constraints can be found in the references above.

A set of benchmark points (BPs) were evaluated in Ref.~\cite{Kalinowski:2018ylg} to identify representative regions of the phase space where the production is still allowed by existing constraints. For the analysis targeting only the same-flavour dilepton final state, the signal cross section is usually dominated by the $AH$ production, and the sensitivity will depend mostly on \mh\ and \mdiff. Instead of just estimating the sensitivity on the benchmark points, we adopt a strategy to scan the parameter space in \mh\ and \mdiff, with several scenarios described below, and summarised in Table~\ref{tab:scens}, to study the dependency on the other parameters: \mch, $\lam_2$ and $\lam_{345}$.

\begin{center}
\begin{table*}[tbh!]
\begin{center}
\begin{tabular}{c|c|c|c}
Scenario & $M_{H^\pm}$ & $\lam_{345}$ & motivation \\
\hline
S1 & $M_A$ & $0$ &targets main contribution process\\
S2 & $M_A$ & $\lam_{max}$ & switches on $h-$Strahlung process\\
S3 & $M_{H^{\pm}}^{max}$ & $\lam_{max}$ &switches on contributions from $H^+\,H^-$ production\\
\end{tabular}
\caption{\label{tab:scens} \vspace{2mm} \\Scenarios considered in this work for the 
scan in the $\lb M_H, M_A-M_H \rb$ plane. Parameters are chosen such that they subsequently add the contributions of the diagrams in figure~\ref{fig:idm_diagram} to the signal process. See text for details.}
\end{center}
\end{table*}
\end{center}

Diagrams involving quartic couplings between IDM scalar bosons do not participate to the production of the final state under study at leading order. Hence, there is no sensitivity to the $\lam_2$ coupling, which is set to 0.1 everywhere. Concerning diagrams involving the coupling $\lam_{345}$ of the new scalars with the SM $h$ (see Fig.~\ref{fig:idm_diagram} bottom) two scenarios are considered: one where $\lam_{345}=0$ and one where $\lam_{345}$ is set to a relatively large value, referred to as $\lam_{max}$, estimated using a linear parametrisation in \mh, $\lam_{max}=0.0018\,\times\,\lb M_H-72\, \GeV \rb$. This parametrization is chosen as an envelope of the parameter points in the $\lb M_H,\,\lam_{345} \rb$ plane that were allowed by early LUX-ZEPLIN results  for direct detection \cite{LZ:2022lsv} and approaches the maximal value for the coupling combination allowed by these constraints. 
As DM bounds in addition depend on the mass of the unstable neutral scalar, \ma, the actual maximal value might be lower than the one given by this parametrisation.

\begin{figure}[h!]
\begin{minipage}{\columnwidth}
       \centering \includegraphics[width=1.\textwidth]{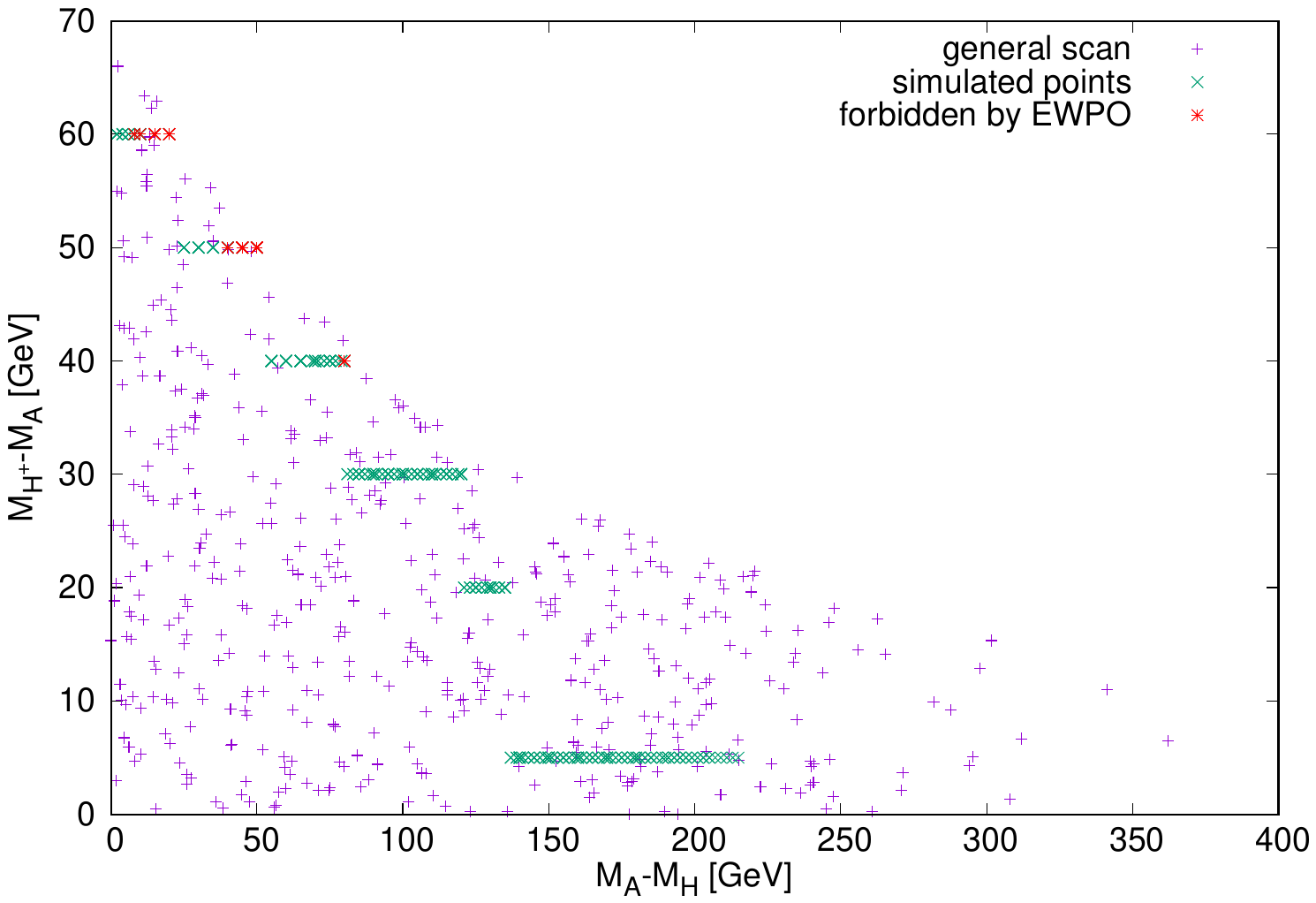}
\end{minipage}
\caption{\label{fig:massdiffscan}  Allowed points in a general scan with $\lam_2\,=\,0.1$ {\slshape (purple, $+$)}, as well as points used in the simulation {\slshape (green, x)}. Some of the latter are forbidden by electroweak precision  observable constraints, which were implemented via the oblique parameters $S,\,T,\,U$ {\slshape (red, $*$)}.
}
\end{figure}

Next, we investigate a scenario that differs in the choice of \mch, from \mch$=$\ma\ to \mch$-$\ma~ set to be within the maximum value allowed, giving a value of \mch\ referred to as $M_{H^{\pm}}^{max}$. Maximal mass splittings are in general mainly constrained from requiring perturbative unitarity as well as electroweak precision data\footnote{We thank J. Braathen for useful discussions regarding this point.}. Note, however, that the latter also depends on the absolute mass scale. In order to estimate the maximally allowed mass differences, we first perform a random scan where we fix $\lam_2$ but let all other parameters float. The resulting allowed parameter space  is shown in figure \ref{fig:massdiffscan}. The points that have been simulated according to the fixed mass differences discussed above are displayed in green. As the absolute scale also plays a role in the determination of the oblique parameters, some points we chose, although overlaying with the allowed regions of a general scan, are forbidden by EWPO for specific choices of mass differences. These are displayed in red, but we chose to keep them for the final result for consistency reasons. Note that $M_{H^\pm}-M_A\neq\,0$ is only realized in one of the three scenarios we consider; therefore, the points that would be forbidden here are not explicitly labeled in our final result plots.

\section{Monte Carlo samples}
\label{sec:mc}

\begin{figure}[h!]
    \begin{minipage}{\columnwidth}
\includegraphics[width=\columnwidth]{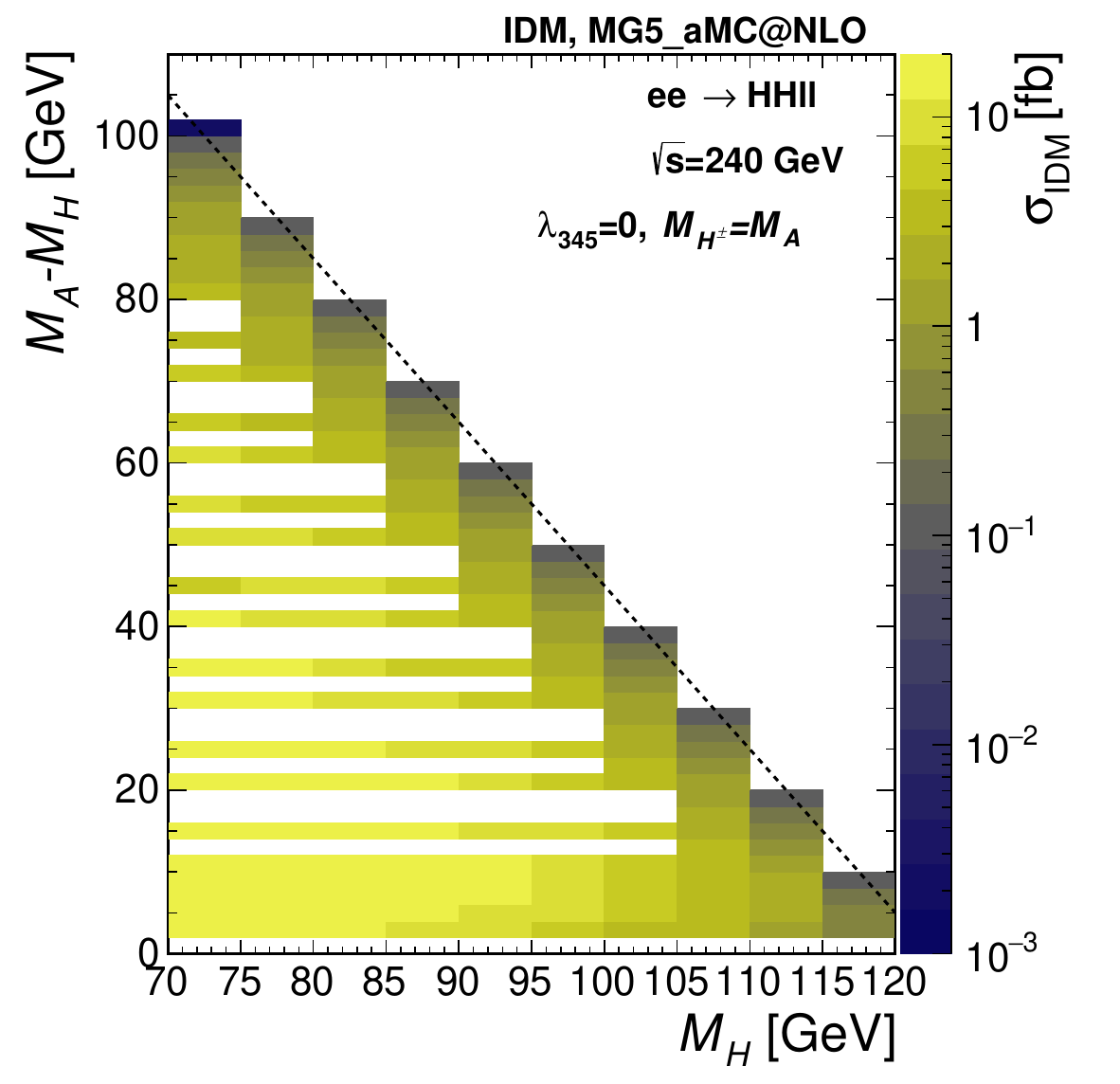}\\
\includegraphics[width=\columnwidth]{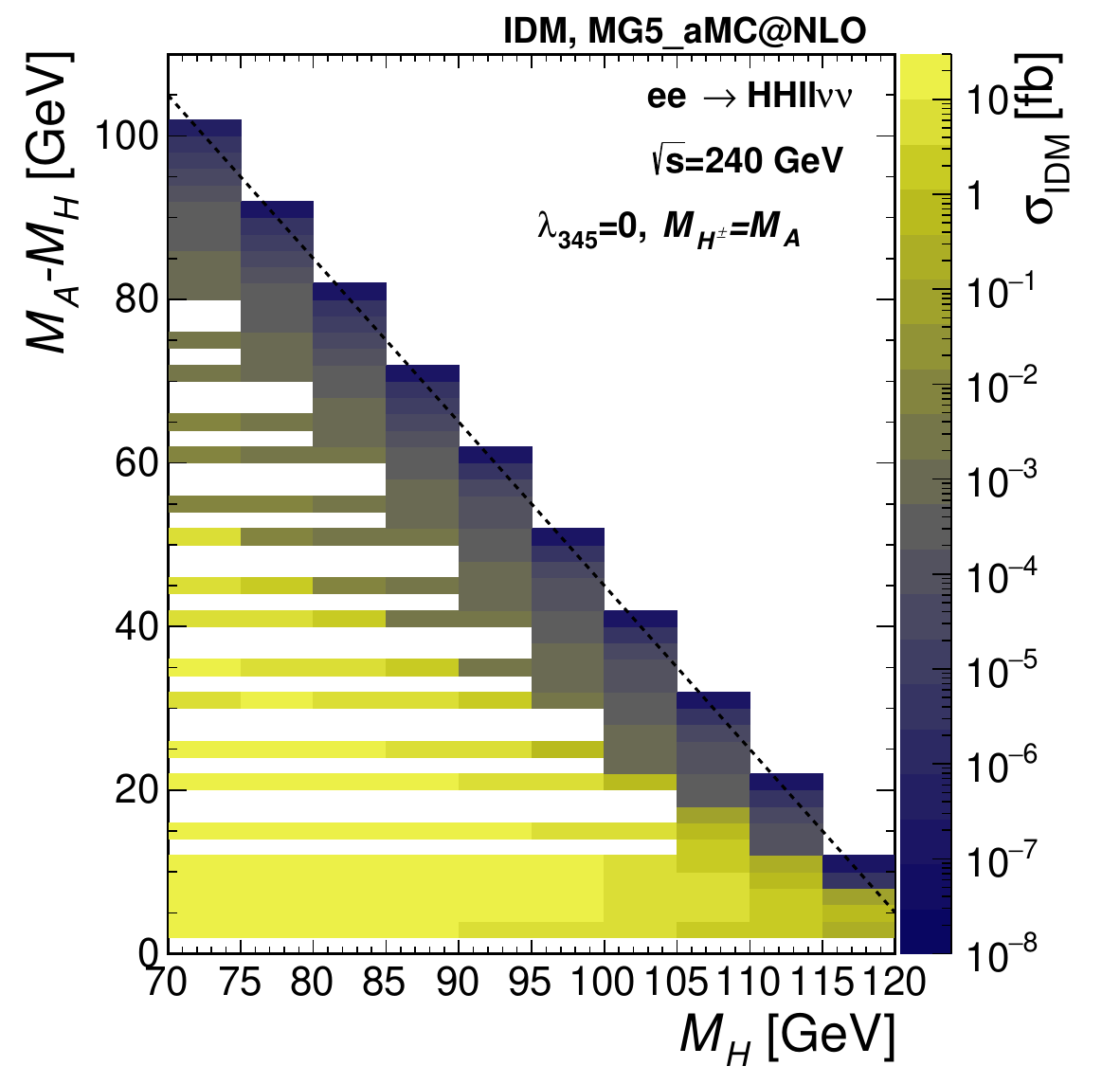}
    \end{minipage}
    \caption{Cross section from the MG5\_aMC@NLO simulation at \sqs{240}, for the points simulated in the \mdiff\ vs \mh\ plane in scenario S1. All cross sections are in fb, computed for $p_{\mathrm{T}}^{\ell}>0.5$\,GeV and with a numerical integration error smaller than 0.1\%. Top: $HH\ell\ell$ final state, bottom: $HH\ell\ell\nu\nu$ final state. The dashed line shows the kinematic limit for on-shell production.}
    \label{fig:grid240}
\end{figure}

\begin{figure*}[h!]
\includegraphics[width=0.45\textwidth]{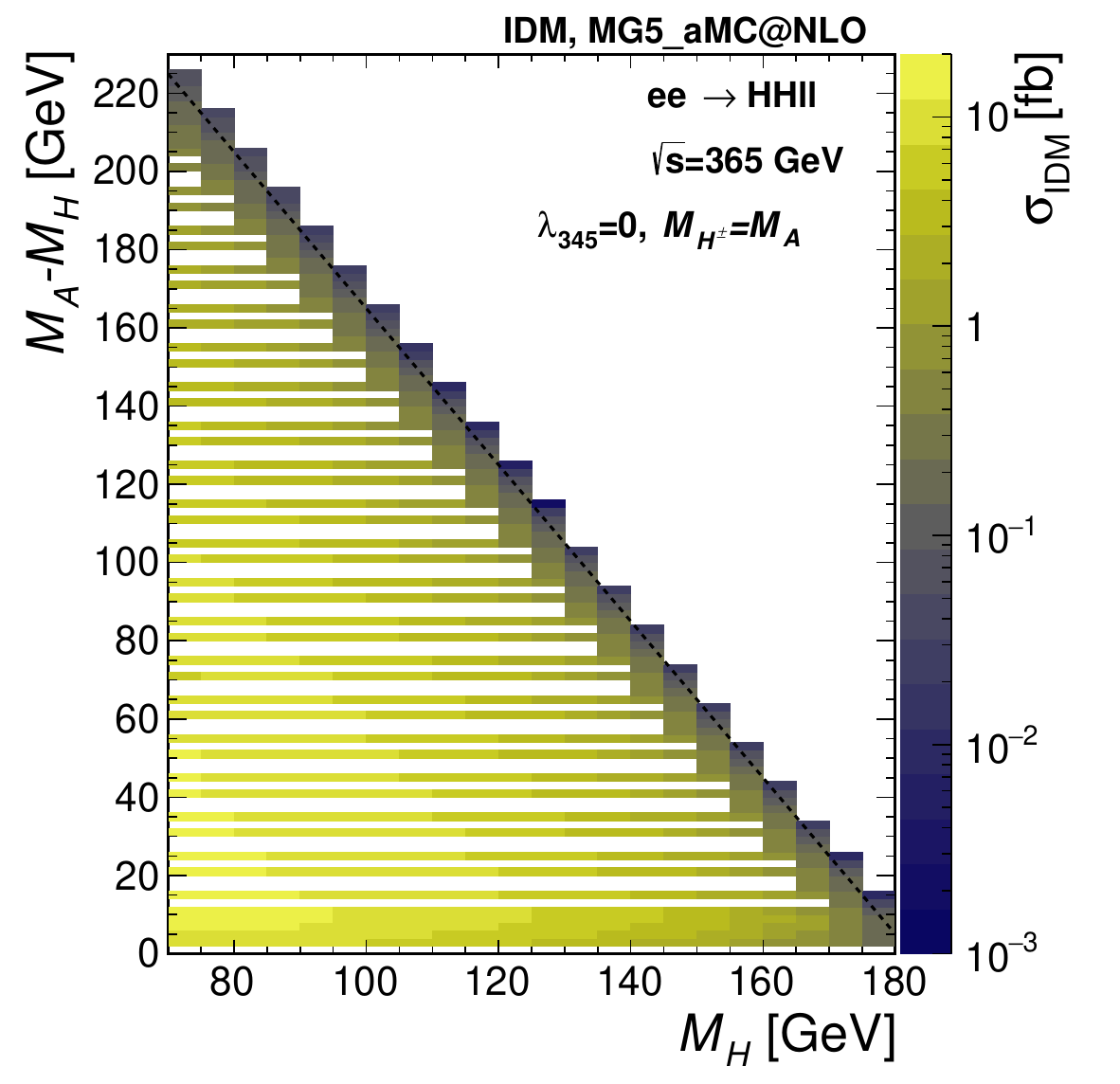}\hfill
\includegraphics[width=0.45\textwidth]{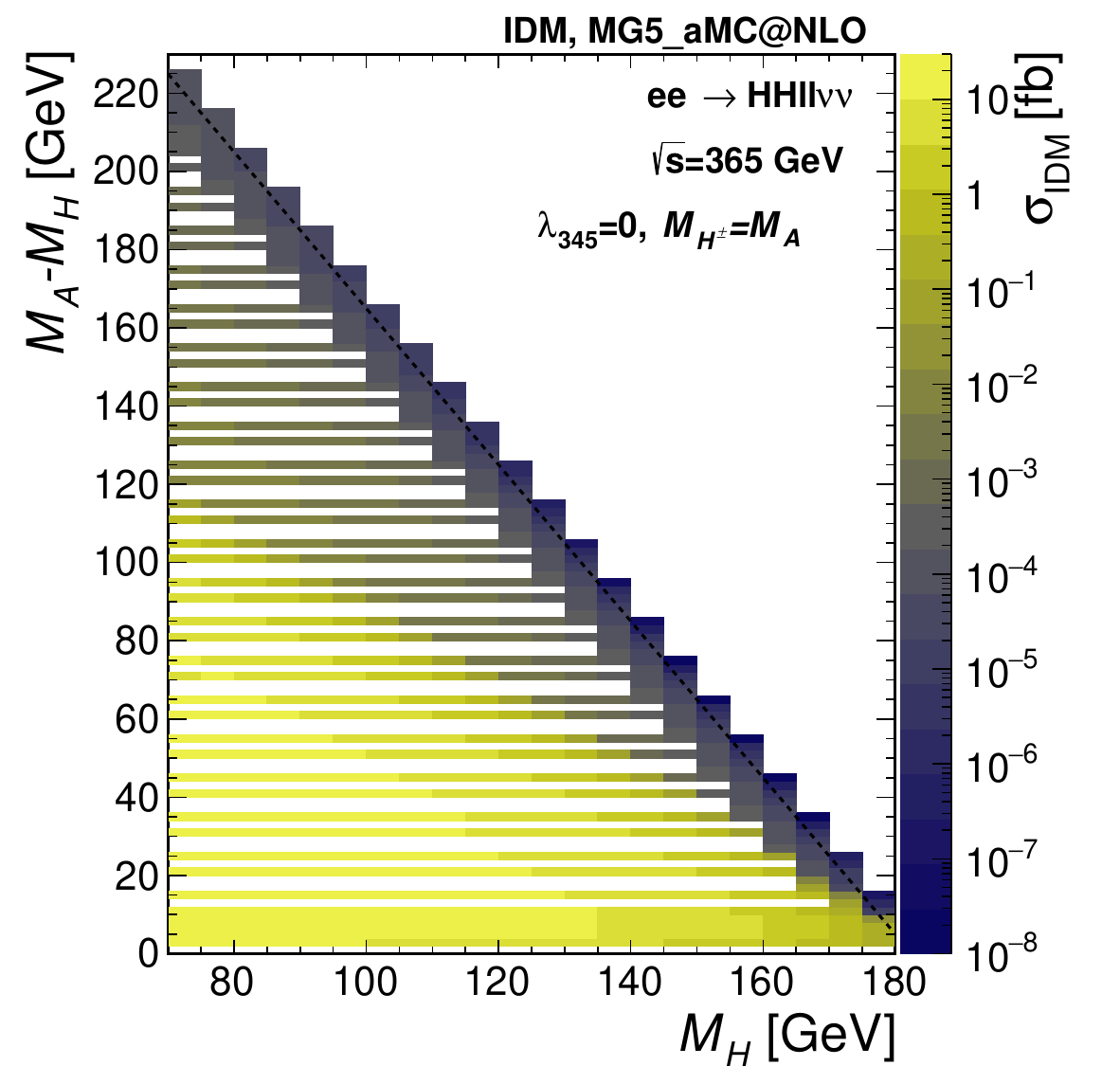}
    \caption{Cross section from the MG5\_aMC@NLO simulation at \sqs{365}, for the points simulated in the \mdiff\ vs \mh\ plane in scenario S1. All cross sections are in fb, computed for $p_{\mathrm{T}}^{\ell}>0.5$ GeV and with a numerical integration accuracy smaller than 0.1\%. Left: $HH\ell\ell$ final state, right: $HH\ell\ell\nu\nu$ final state. The dashed line shows the kinematic limit for on-shell production.}
    \label{fig:grid365}
\end{figure*}

The IDM signal samples are generated using MADGRAPH5\_aMC@NLO v2.8.1~\cite{Alwall:2014hca} interfaced with PYTHIA v8.2~\cite{Sjostrand:2014zea}, using $e^+e^-$ collisions at \sqs{240 and 365}. The input UFO~\cite{Degrande:2011ua} model is taken from Ref.~\cite{Goudelis:2013uca}. Instead of targeting individual pair production modes, two final states are generated directly, therefore taking into account all contributing diagrams and the proper interferences. The two final states we consider here are given by $HH\ell\ell$ and $HH\ell\ell\nu\nu$, with $\ell=e$, $\mu$ or $\tau$, and $H$ the stable dark scalar\footnote{Here and in the following, we use the shorthand notation $\ell\ell\,\equiv\,\ell^+\,\ell^-$ and $\nu\,\nu\,\equiv\,\nu\,\bar{\nu}$. In all processes discussed here, both electroweak charge and lepton number are conserved.}. The lepton charges are omitted in the notations. The $\tau$ leptons are decayed by PYTHIA, and the contribution from the $\tau$ leptonic decay modes leading to two electrons or two muons are included in the signal. Typically, the former final state is obtained with a virtual or real $Z$ boson (Fig.~\ref{fig:idm_diagram} top), and the sensitivity will roughly depend on the mass difference between $H$ and $A$. The latter final state is mediated via a pair of $W$ bosons (Fig.~\ref{fig:idm_diagram} middle) and the sensitivity will instead depend on the difference in mass between $H$ and the charged scalars. The signal is generated fixing the coupling $\lambda_2$ to 0.1, \mh\ between 70 and 115 (180) GeV in steps of 5 GeV for $\sqs{240\,(365)}$, and \mdiff\ between 2 GeV and the kinematic limit for on-shell production in steps of 2 to 5 GeV. Three scenarios are considered for setting \mch\ and $\lam_{345}$, as detailed in section~\ref{sec:idm}, and 500,000 events are generated per point. The cross section values are shown in Figs.~\ref{fig:grid240} (\ref{fig:grid365}) in the simulated \mdiff\ vs. \mh\ grid plane for \sqs{240 (365)}, on the top (left) for the $HH\ell\ell$ final state and on the bottom (right) for the $HH\ell\ell\nu\nu$ final state, for the scenario S1.  For the number of events generated, the Monte Carlo (MC) integration error for the cross sections is typically smaller than 0.1\%. The minimum transverse momentum of the leptons, $p_{\mathrm{T}}^{\ell}$, is set at 0.5 GeV.

\begin{figure*}[h!]
\includegraphics[width=0.45\textwidth]{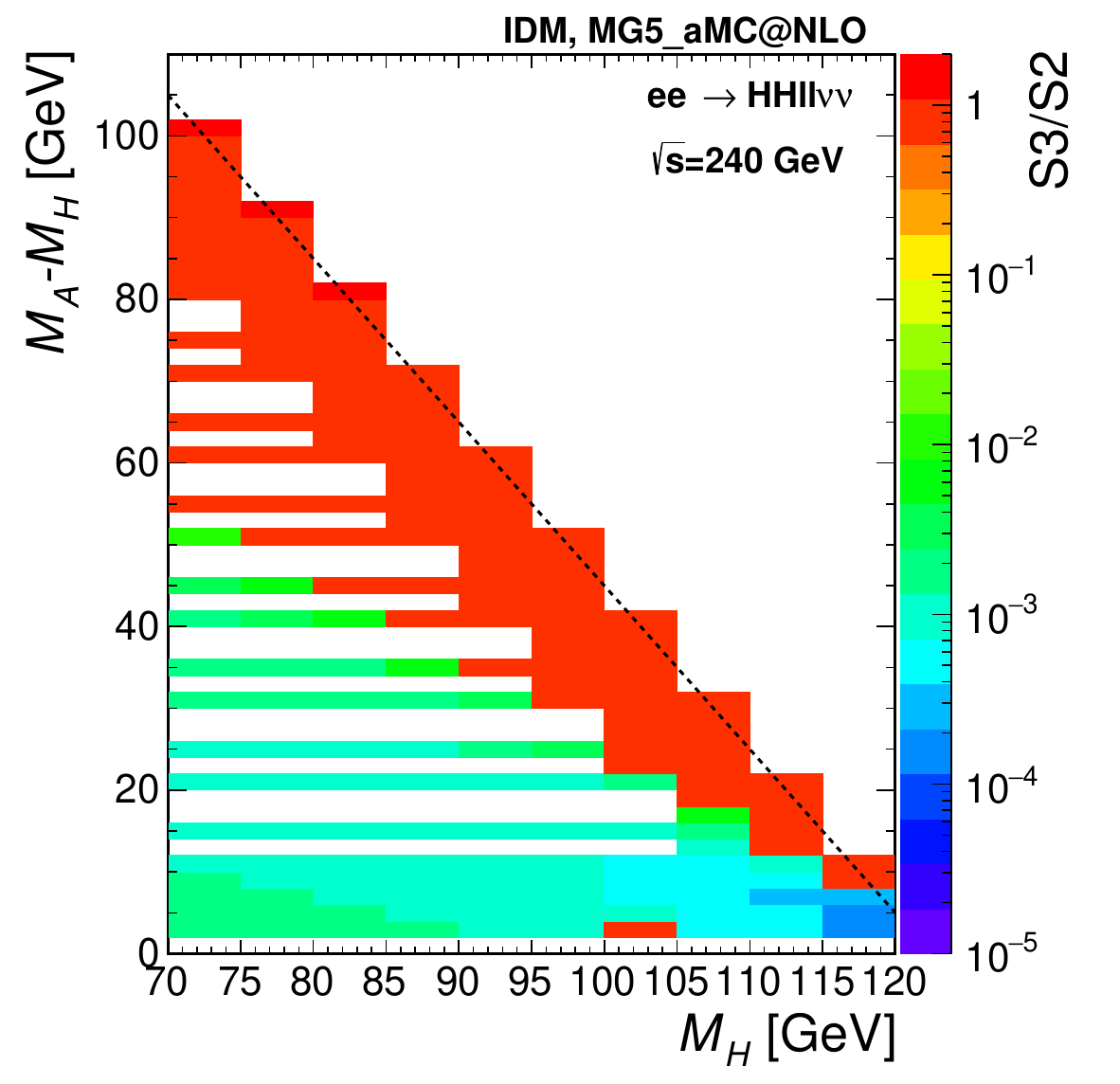}\hfill
\includegraphics[width=0.45\textwidth]{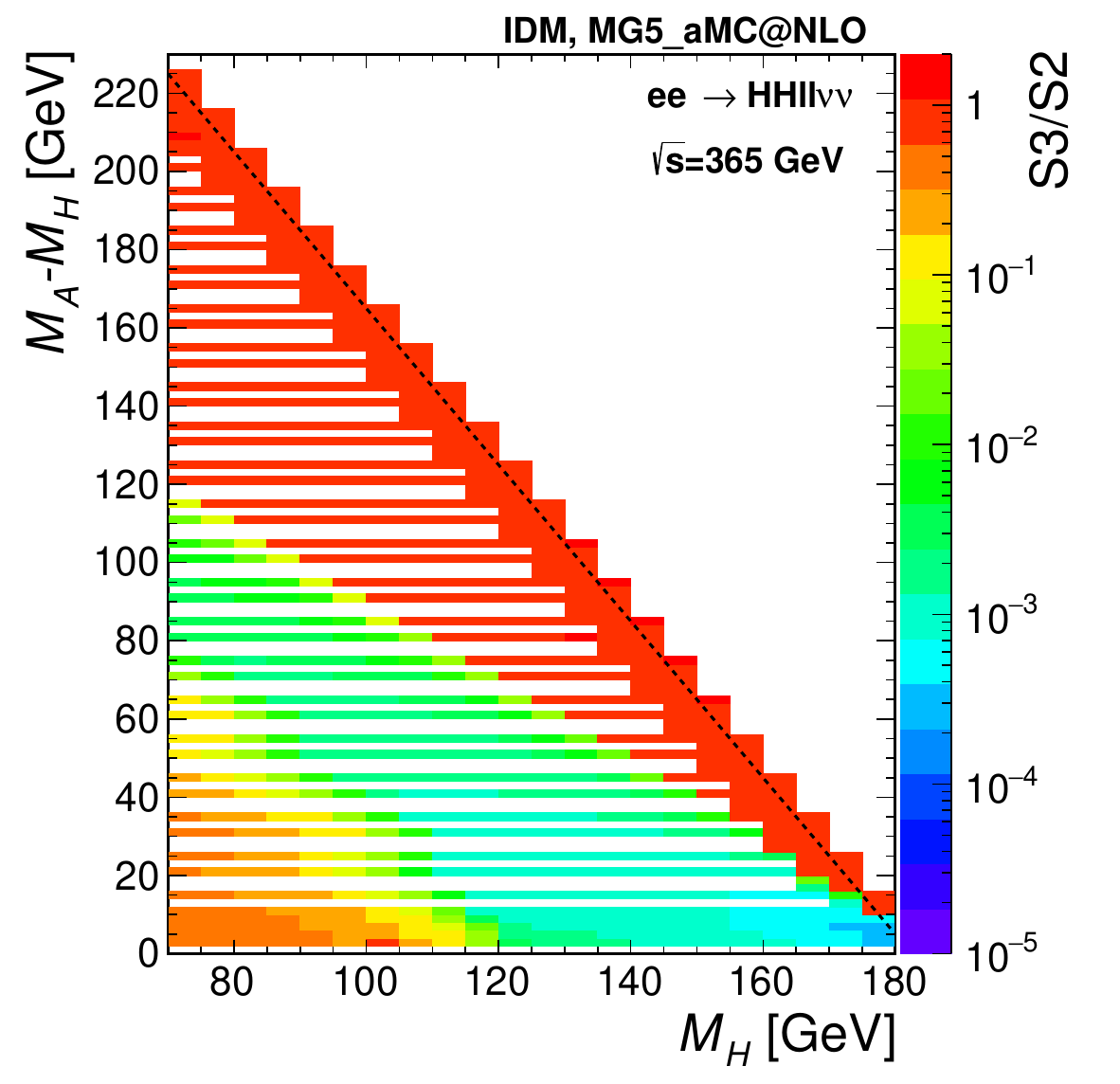}
    \caption{Cross section ratios S3/S2 in the \mdiff\ vs \mh\ plane, from the MG5\_aMC@NLO simulation at \sqs{240} (left) and \sqs{365} (right), for the $HH\ell\ell\nu\nu$ final state. The dashed line shows the kinematic limit.}
    \label{fig:S3S2}
\end{figure*}

Almost no difference is found between S2 and S1, which is expected from the range of \mh\ values of interest here, \mh$>$\msmh$/2$, leading to the suppression of the BSM decay of $h$ into $HH$. The largest differences are for S3 in the $HH\ell\ell\nu\nu$ final state, for which \mch\ drives the cross section. The cross section ratios between S3 and S2 are shown in  Fig.~\ref{fig:S3S2} in the \mdiff\ vs. \mh\ plane for \sqs{240} (left) and \sqs{365} (right) for the $HH\ell\ell\nu\nu$ final state.

\clearpage
Background processes include diboson production ($WW$ and $ZZ$), inclusive $ee$, $\mu\mu$ and $\tau\tau$ production, SM Higgs boson production in association with a $Z$ boson, and \mytt\ production at \sqs{365}. Samples were produced centrally with high statistics, within the so-called Winter2023 campaign~\cite{winter23}, as given in Table~\ref{tab:bkg240} for \sqs{240} and in Table~\ref{tab:bkg365} for \sqs{365}, using PYTHIA v8.306~\cite{Bierlich:2022pfr} for the diboson and \mytt\ samples, and WHIZARD v3.0.3~\cite{Kilian:2007gr} interfaced with PYTHIA v6.4.28~\cite{Sjostrand:2006za} for the other processes. For the inclusive $ee$ production, a lower limit of 30 GeV is set on the invariant mass of the pair of electrons produced. The dielectron channel is hence included in the analysis only in the parameter space where the background is simulated, namely for \mdiff$>30$\,GeV. 

The current target values for the total integrated luminosities of the \sqs{240 and 365} FCC-ee runs are 10.8 and 2.7~\ab, respectively~\cite{janot_2024_nfs96-89q08}.

Both signal and background samples are processed through DELPHES v3.5.1pre05~\cite{deFavereau:2013fsa} with the IDEA~\cite{Tassielli:2021rjk} detector configuration as per the Winter2023 campaign. Events are produced using the KEY4HEP~\cite{Ganis:2021vgv} framework and analysed using the FCCAnalyses package~\cite{fccana}.

\begin{table*}[h!]
    \centering
  \begin{tabular}{l|c|c|c}
Process & N generated & cross section (pb) & Eq. $\mathcal{L}$ (ab$^{-1}$) \\
\hline
 $ZZ$      &	56162093  & 1.359   & 41\\
 $WW$      &	373375386 & 16.4385 & 23\\
$Z${($ee$)$h$}    &	1200000	  & 0.00716 & 168\\
$Z${($\mu\mu$)$h$}   &	1200000   & 0.00676 & 178\\
$Z${($\tau\tau$)$h$} &	1200000   & 0.00675 & 178\\
$Z${($\nu\nu$)$h$}   &	3500000	  & 0.0462  & 76\\
$Z${($\mathrm{q}\bar{\mathrm{q}}$)$h$}     &	6700000	  & 0.136   & 49\\
$ee$ $30<M_{ee} < 150$\,\GeV &	85400000  & 8.305   & 10\\
$\mu\mu$    &	53400000  & 5.288   & 10\\
$\tau\tau$  &	52400000  & 4.668   & 11\\
  \end{tabular}
    \caption{Background processes generated in the Winter2023 campaign at \sqs{240}.  $ZZ$ and $WW$ are generated using Pythia8, whilst the others are generated with Whizard+Pythia6. The last column gives the total integrated luminosity simulated.}
    \label{tab:bkg240}
\end{table*}
\begin{table*}[h!]
    \centering
  \begin{tabular}{l|c|c|c}
Process & N generated & cross section (pb) & Eq. $\mathcal{L}$ (ab$^{-1}$) \\
\hline
$ZZ$	& 11470944	& 0.6428	& 18 \\
 $WW$	& 11754213	& 10.7165	& 1.1 \\
$Z${($ee$)$h$}	& 1000000	& 0.00739	& 135\\
$Z${($\mu\mu$)$h$}	& 1200000	& 0.004185	& 287 \\
$Z${($\tau\tau$)$h$}	& 1100000	& 0.004172	& 264 \\
$Z${($\nu\nu$)$h$}	& 2200000	& 0.05394	& 41 \\
$Z${($\mathrm{q}\bar{\mathrm{q}}$)$h$}	& 2400000	& 0.032997	& 73 \\
$ee$ $30<M_{ee} < 150$\,\GeV	& 3000000	& 1.527	& 2.0 \\
$\mu\mu$	& 6600000	& 2.2858	& 2.9 \\
$\tau\tau$	& 12800000	& 2.01656	& 6.3 \\
\mytt	& 2700000	& 0.8	& 3.4 \\
  \end{tabular}
    \caption{Background processes generated in the Winter2023 campaign at \sqs{365}.  $ZZ$,  $WW$ and  $t\bar{t}$ are generated using Pythia8, whilst the others are generated with Whizard+Pythia6. The last column gives the total integrated luminosity simulated.
    }
    \label{tab:bkg365}
\end{table*}
\section{Definition of the objects in Delphes}
\label{sec:setup}

The origin of the coordinate system adopted in this work is defined as having the origin centered at the nominal collision point inside the experiment. The y-axis points vertically upward, and the x-axis points radially inward toward the centre of the FCC. Thus, the z-axis points along the beam direction. The azimuthal angle $\phi$ is measured from the x-axis in the x-y plane. The polar angle $\theta$ is measured from the z-axis. Pseudorapidity is defined as $\eta=-\ln\left[ \tan(\theta /2)\right]$. The momentum and energy are denoted by $p$ and $E$, respectively. Those transverse to the beam direction, denoted by \pT and \ET, respectively, are computed from the x and y components. The imbalance of energy is denoted by \emiss, and that measured in the transverse plane by \etmiss.

The parametrisation of the reconstructed objects in Delphes is summarised in Table~\ref{tab:delphes}. The resolution smearing for the calorimeters is defined as 
$$\sigma_E = \sqrt{E^2\times c^2 + E \times s^2 + n^2}$$
with $c=0.005$ (0.01) the constant term, $s=0.03$ (0.3)\,GeV$^{1/2}$ the sampling term and $n=0.002$ (0.05)\,GeV the noise term for the electromagnetic (hadronic) parts.

At preselection, a minimum momentum requirement of $p>5$\,GeV is imposed on the Delphes reconstructed electrons, muons and photons. Events with exactly two such electrons or exactly two such muons are selected for analysis.

Jets are reclustered from Delphes reconstructed particles after removing the selected electron and muon candidates. The Durham algorithm is used, with exclusive clustering of two jets, and an energy-based scheme~\cite{Cacciari:2011ma}, which leads to good resolution in low-multiplicity final states, and is a standard choice for lepton colliders. The missing energy is taken as given by Delphes, where the DM candidates count as invisible particles. 


\begin{table*}[h]
    \centering
    \begin{tabular}{l|c|c|c}
    Algorithm & Objects & Selection requirements & efficiency \\
    \hline
       Tracking  & $e$, $\mu$, charged hadrons & \pT$>0.1$\,GeV, $|\eta|<2.56$ & 1 \\
       Identification & $\gamma$, $e$, $\mu$ & $E>2$\,GeV, $|\eta|<3$ & 0.99 \\ 
    \end{tabular}
    \caption{Efficiency and selection requirements implemented in the Delphes card for the IDEA detector configuration.
    }
    \label{tab:delphes}
\end{table*}


\section{Background rejection}
\label{sec:sel}

To remain fully sensitive to all signal points but significantly reduce the background contributions, the following preselection criteria are applied, which are inspired from previous studies performed at generator level with the CLIC setup~\cite{Kalinowski:2018kdn,Zarnecki:2020swm}.

\begin{figure}[h!]
    \begin{minipage}{\columnwidth}
\includegraphics[width=\textwidth]{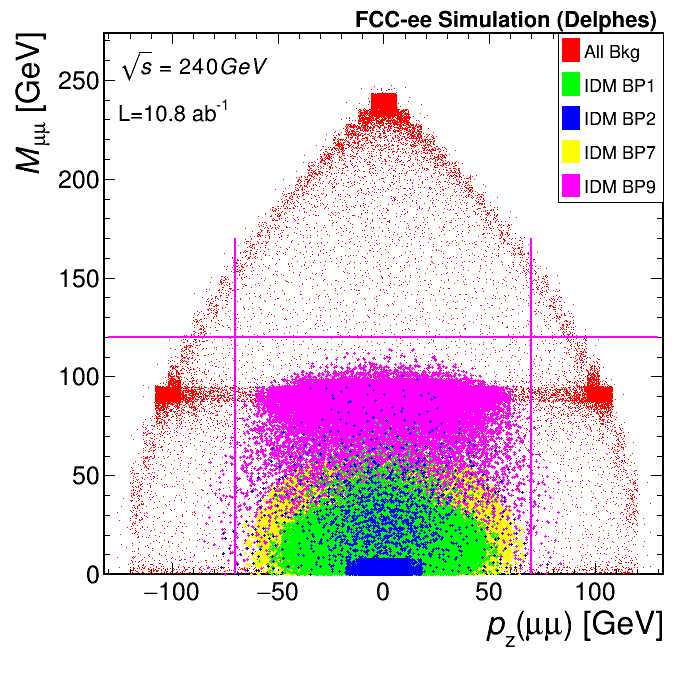}\\
\includegraphics[width=\textwidth]{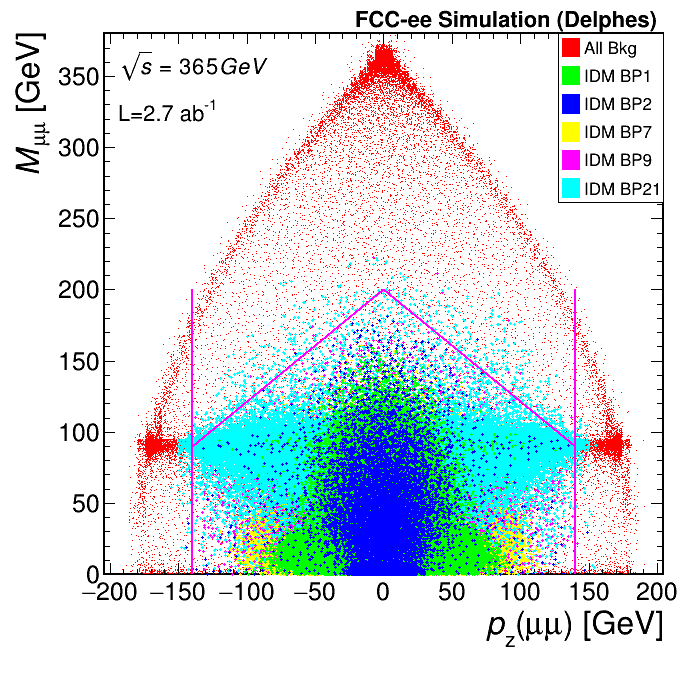}
\end{minipage}
    \caption{Dimuon mass $M_{\mu\mu}$ as a function of the dimuon $p_z(\mu\mu)$, for the selection of a muon pair with $e^+e^-$ collisions at \sqs{240 (365)} on the top (bottom). The sum of the background processes is shown in red. Several representative signal benchmark points are shown in green (BP1), blue (BP2), yellow (BP7) and pink (BP9). On the bottom, also BP21 is shown in cyan. The specification of the BPs can be found in Ref.~\cite{Kalinowski:2018kdn}. The pink lines highlight the selection requirements.
    }
    \label{fig:2dfcc}
\end{figure}

Only events with exactly two same-flavour leptons $\ell=e$ or $\mu$ (including decays of $\tau$ leptons to $e$ or $\mu$), $M_{\ell\ell}<120$\,GeV, $|p_z(\ell\ell)|<70$\,GeV are kept. For the \sqs{365} setup, for several benchmark points the signal contribution has $M_{\ell\ell}$ values well above the Z invariant mass. In order to more efficiently reject the backgrounds, the selection is modified to $|p_z(\ell\ell)|<140$\,GeV, and the selection in $M_{\ell\ell}$ is chosen to be dependent on $p_z(\ell\ell)$, with the formula: $M_{\ell\ell}<(-9.0/14.0 \times |p_z(\ell\ell)| + 200)$, with $M_{\ell\ell}$ and $p_z(\ell\ell)$ in GeV. As an example, the distribution of $M_{\mu\mu}$ as a function of $p_z(\mu\mu)$ is shown for the sum of the major backgrounds and a few signal representative benchmark points from Ref.~\cite{Kalinowski:2018kdn} in Fig.~\ref{fig:2dfcc}, for \sqs{240 (365)} on the top (bottom). A good agreement is found with the results obtained using the CLIC setup.

With a negligible impact on the signal efficiency, the $ee\rightarrow ee,\,\mu\mu,\,\tau\tau$ background is further reduced by 90\% if requiring the transverse missing energy \etmiss$>5$\,GeV. As no additional object is expected for signal events, a veto is applied on events with any additional electron or muon, any photon  with $E>5\,\GeV$ or any jet. To further reject background, only events with the leading (subleading) lepton \pT$<80$ (60)\,GeV are kept for \sqs{240}, \pT$<140$ (80)\,GeV for \sqs{365}. Finally, events with $p_{\ell\ell}/E_{\ell\ell}>0.1$ are selected. All selection criteria are summarised in Table~\ref{tab:sel}.

\begin{table*}[h!]
    \centering
    \begin{tabular}{l|c|c|c}
       Step  & Selection at \sqs{240} & Selection at \sqs{365} & target background\\
       \hline
    Preselection & $|p_z(\ell\ell)|<70$\,GeV &  $|p_z(\ell\ell)|<140$\,GeV & $ZZ$, $ee\rightarrow ee,\,\mu\mu$ \\
     & $M_{\ell\ell}<120$\,GeV & $M_{\ell\ell}<(-9.0/14.0 \times |p_z(\ell\ell)| + 200)$ & $WW$, $ee\rightarrow ee,\,\mu\mu$\\
     & \multicolumn{2}{c|}{\etmiss$>5$\,GeV} & $ZZ$, $ee\rightarrow ee,\,\mu\mu$ \\
     \hline
     Object veto & \multicolumn{2}{c|}{3$^{\mathrm{rd}}$ lepton  $p>5\,\GeV$}, jet, photon $E>5$\,GeV & $WW$, $ZZ$, $ee\rightarrow ee,\,\mu\mu,\,\tau\tau$ \\
     Leptons \pT &  \pT$<80,60$\,GeV & \pT$<140,80$\,GeV & $WW$, $ZZ$, $ee\rightarrow\tau\tau$ \\
     E/p & \multicolumn{2}{c|}{$p_{\ell\ell}/E_{\ell\ell}>0.1$} & $ee\rightarrow ee,\,\mu\mu,\,\tau\tau$ \\
    \end{tabular}
    \caption{Summary of the selection criteria and target backgrounds. Here $\ell$ is used for either electron or muon.}
    \label{tab:sel}
\end{table*}

Just from requiring a pair of same-flavour leptons, 16\% (4\%) of $ZZ$ ($WW$) events are selected, further reduced to 5\% (2\%) after preselection, and 1.5\% (1\%) after the veto on other objects and final set of selection criteria. For the dilepton production, the preselection achieves 99\% (94\%) rejection of $ee$ events in the $ee$ channel for \sqs{240 (365)}, 99\% for the $\mu\mu$ in $\mu\mu$ channels, and 97\%  for the $\tau\tau$ events in both channels. The object veto and other selection criteria reduce the $ee$, $\mu\mu$ and $\tau\tau$ backgrounds by another factor of about 40\%, 70\% and 35\%, respectively. For the signal, after requiring a pair of same-flavour leptons with $p>5$\,\GeV\, only 2\% (0.1\%) of events are selected at low mass splitting \mdiff, increasing to about 65\% (6\%) at higher mass splitting, for the production of $HH\ell\ell$ ($HH\ell\ell\nu\nu$) in the scenario S1.

The final number of events expected after selection are given in Table~\ref{tab:Nevts} with the MC statistical uncertainty, for the different background processes and several representative signal points from the scenario S1, normalised to the total integrated luminosity values of the FCC-ee runs at \sqs{240 and 365}, namely 10.8 and 2.7~\ab.

\begin{table*}[h!]
    \centering
    \resizebox{\textwidth}{!}{
        \begin{tabular}{l||c|c||c|c} 
         Process & $ee$ \sqs{240} & $\mu\mu$ \sqs{240}  & $ee$ \sqs{365} & $\mu\mu$ \sqs{365} \\
        \hline
        \multicolumn{5}{c}{Background processes}\\
        \hline
         $ZZ$  & 9.74e+04 $\pm$ 1.60e+02 & 1.09e+05 $\pm$ 1.69e+02& 9.30e+03 $\pm$ 3.75e+01 & 1.05e+04 $\pm$ 3.98e+01 \\
         $WW$  & 8.02e+05 $\pm$ 6.18e+02 & 8.72e+05 $\pm$ 6.44e+02 &  8.75e+04 $\pm$ 4.64e+02 & 9.32e+04 $\pm$ 4.79e+02 \\ 
        \mytt & - & - & 0 & 0 \\
        $ee$   & 6.16e+05 $\pm$ 8.05e+02 & 0  & 1.34e+05 $\pm$ 4.30e+02 & 0 \\
        $\mu\mu$  & 0 & 9.80e+04 $\pm$ 3.24e+02 & 0 & 2.86e+04 $\pm$ 1.64e+02 \\
        $\tau\tau$  & 4.92e+05 $\pm$ 6.88e+02 & 4.80e+05 $\pm$ 6.80e+02& 5.83e+04 $\pm$ 1.57e+02 & 5.65e+04 $\pm$ 1.55e+02 \\
         $Z${($ee$)$h$}  & 1.11e+02 $\pm$ 2.68e+00 & 0 &  1.88e+01 $\pm$ 6.13e-01 & 9.98e-02 $\pm$ 4.46e-02 \\
         $Z${($\mu\mu$)$h$}  & 6.09e-02 $\pm$ 6.09e-02 & 1.33e+02 $\pm$ 2.85e+00&  9.42e-03 $\pm$ 9.42e-03 & 1.74e+01 $\pm$ 4.05e-01 \\
        $Z${($\nu\nu$)$h$}  & 2.15e+03 $\pm$ 1.75e+01 & 2.25e+03 $\pm$ 1.79e+01 & 6.08e+02 $\pm$ 6.34e+00 & 6.72e+02 $\pm$ 6.67e+00 \\
        $Z${($\tau\tau$)$h$}  & 1.62e+01 $\pm$ 9.93e-01 & 1.90e+01 $\pm$ 1.08e+00 &2.41e+00 $\pm$ 1.57e-01 & 2.11e+00 $\pm$ 1.47e-01 \\
        $Z${(\myqq)$h$}  & 0 & 0 & 0& 0 \\
        \hline
                \hline
        \multicolumn{5}{c}{Representative signal mass points} \\
        \hline
        70-76 $HH\ell\ell$ & 6.31e+03 $\pm$ 4.16e+01 & 7.07e+03 $\pm$ 4.41e+01  & - & - \\
        70-76 $HH\ell\ell\nu\nu$ & 1.24e+03 $\pm$ 2.33e+01 & 1.45e+03 $\pm$ 2.53e+01  & - & - \\
        80-150 $HH\ell\ell$ & 1.67e+03 $\pm$ 4.55e+00 & 1.86e+03 $\pm$ 4.80e+00  & - & - \\
        80-150 $HH\ell\ell\nu\nu$ & 3.06e-05 $\pm$ 7.42e-06 & 3.96e-05 $\pm$ 8.44e-06  & - & - \\
        100-134 $HH\ell\ell$ & 8.18e+02 $\pm$ 2.25e+00 & 8.94e+02 $\pm$ 2.35e+00  & - & - \\
        100-134 $HH\ell\ell\nu\nu$ & 1.14e-06 $\pm$ 6.59e-07 & 3.81e-07 $\pm$ 3.81e-07  & - & - \\
        \hline
        100-104 $HH\ell\ell$ & - & - & 1.09e+02 $\pm$ 1.59e+00 & 1.39e+02 $\pm$ 1.80e+00 \\
        100-104 $HH\ell\ell\nu\nu$ & - & - & 1.96e+01 $\pm$ 1.00e+00 & 1.96e+01 $\pm$ 1.00e+00 \\
        120-231 $HH\ell\ell$ & - & - & 1.25e+02 $\pm$ 3.47e-01 & 1.41e+02 $\pm$ 3.68e-01 \\
        120-231 $HH\ell\ell\nu\nu$ & - & - & 2.61e-04 $\pm$ 7.59e-06 & 2.07e-04 $\pm$ 6.77e-06 \\
        125-215 $HH\ell\ell$ & - & - & 3.14e+02 $\pm$ 8.65e-01 & 3.53e+02 $\pm$ 9.18e-01 \\
        125-215 $HH\ell\ell\nu\nu$ & - & - & 1.23e-02 $\pm$ 1.14e-04 & 9.92e-03 $\pm$ 1.03e-04 \\
        140-200 $HH\ell\ell$ & - & - & 3.42e+02 $\pm$ 9.29e-01 & 3.80e+02 $\pm$ 9.79e-01 \\
        140-200 $HH\ell\ell\nu\nu$ & - & - & 1.31e-02 $\pm$ 1.21e-04 & 1.06e-02 $\pm$ 1.09e-04 \\
        160-185 $HH\ell\ell$ & - & - & 2.19e+02 $\pm$ 6.29e-01 & 2.39e+02 $\pm$ 6.56e-01 \\
        160-185 $HH\ell\ell\nu\nu$  & - & - & 3.61e-03 $\pm$ 4.81e-05 & 2.83e-03 $\pm$ 4.26e-05 \\
    \end{tabular}} 
    \caption{Expected number of events after selection for both channels ($ee$ and $\mu\mu$) normalised to the total integrated luminosity values of the FCC-ee runs at \sqs{240 and 365}, namely 10.8 and 2.7~\ab. The signal points are all shown for scenario S1, labelled using the simulated values of \mh\ and \ma\ given in GeV in that order, and separately for the two generated final states, $HH\ell\ell$ and $HH\ell\ell\nu\nu$. Different representative mass points are chosen for the different centre-of-mass scenarios. Only statistical uncertainties are included.
    } 
    \label{tab:Nevts} 
\end{table*}


\section{Search strategy}
\label{sec:pnn}

To maximise the sensitivity and to fully exploit the kinematic differences between the IDM and the SM backgrounds, a Neural Network (NN) is used to separate signal from background after the baseline selections. The kinematic features used as input into the model are similar to those used in Refs.~\cite{Kalinowski:2018kdn,Zarnecki:2020swm} which include both low-level and high-level, derived, event features.

The input variables are: 
\begin{itemize}
\item the dilepton pair energy and momenta $E_{\ell\ell}$, $p_{\mathrm{T}}^{\ell\ell}$, $p_z^{\ell\ell}$,
\item the dilepton invariant mass $M_{\ell\ell}$, 
\item the dilepton recoil mass calculated assuming the nominal $\sqrt{s}$,  
\item the dilepton Lorentz boost $p_{\ell\ell}/E_{\ell\ell}$, 
\item the polar angle of the dilepton pair cos$\theta$, 
\item the leptons \pT and cos($\Delta\phi$), with $\Delta\phi$ the difference in $\phi$ between the two selected leptons,
\item $\ell^-$ production angle with respect to the  beam direction calculated in the dilepton centre-of-mass frame cos($\theta^*$),
\item $\ell^-$ production angle with respect to the dilepton pair boost direction, calculated in the dilepton centre-of-mass frame cos($\theta_R$)
\end{itemize}

\begin{figure*}[h!]
\includegraphics[width=0.45\textwidth]{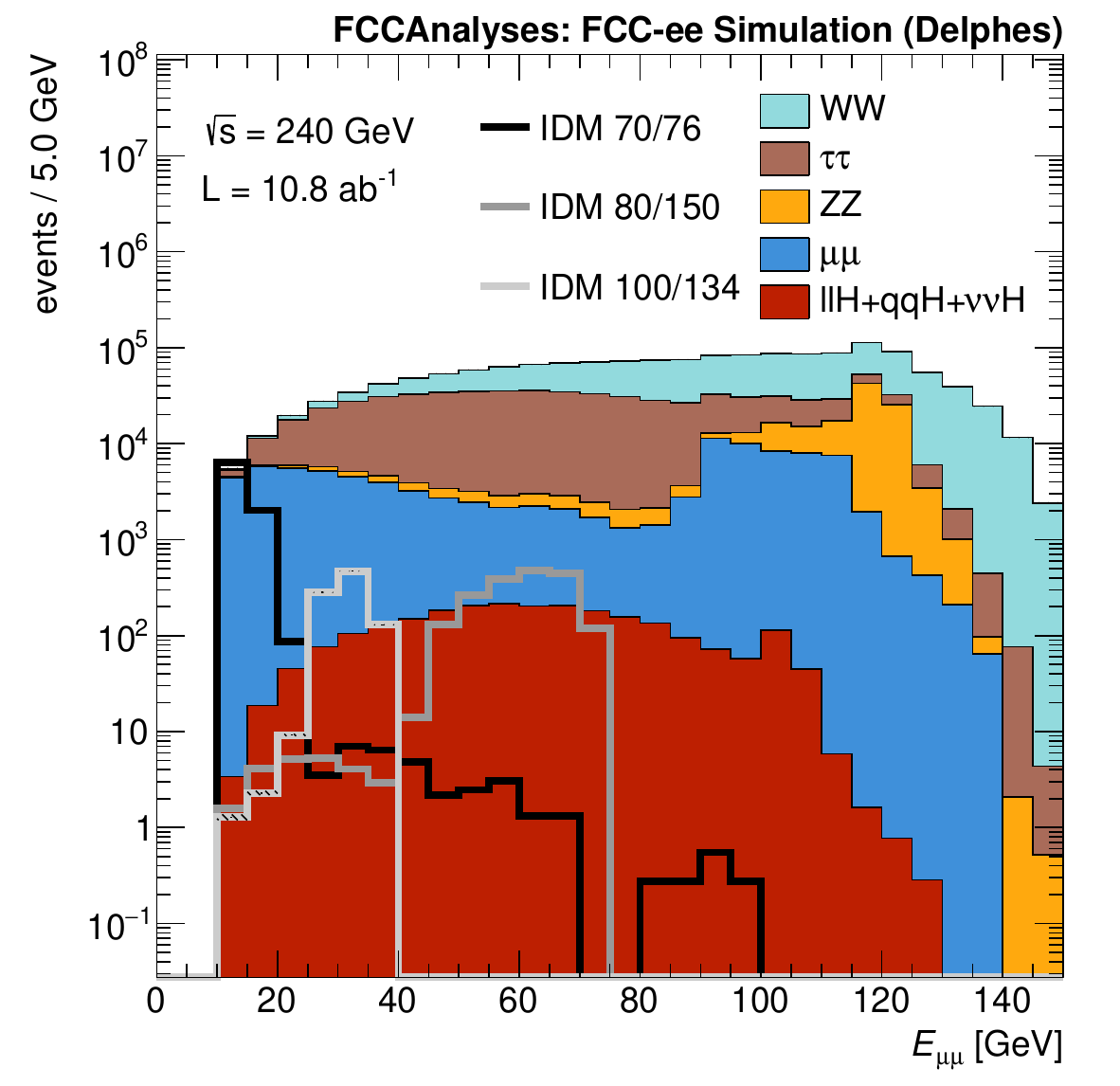}
\hfill
\includegraphics[width=0.45\textwidth]{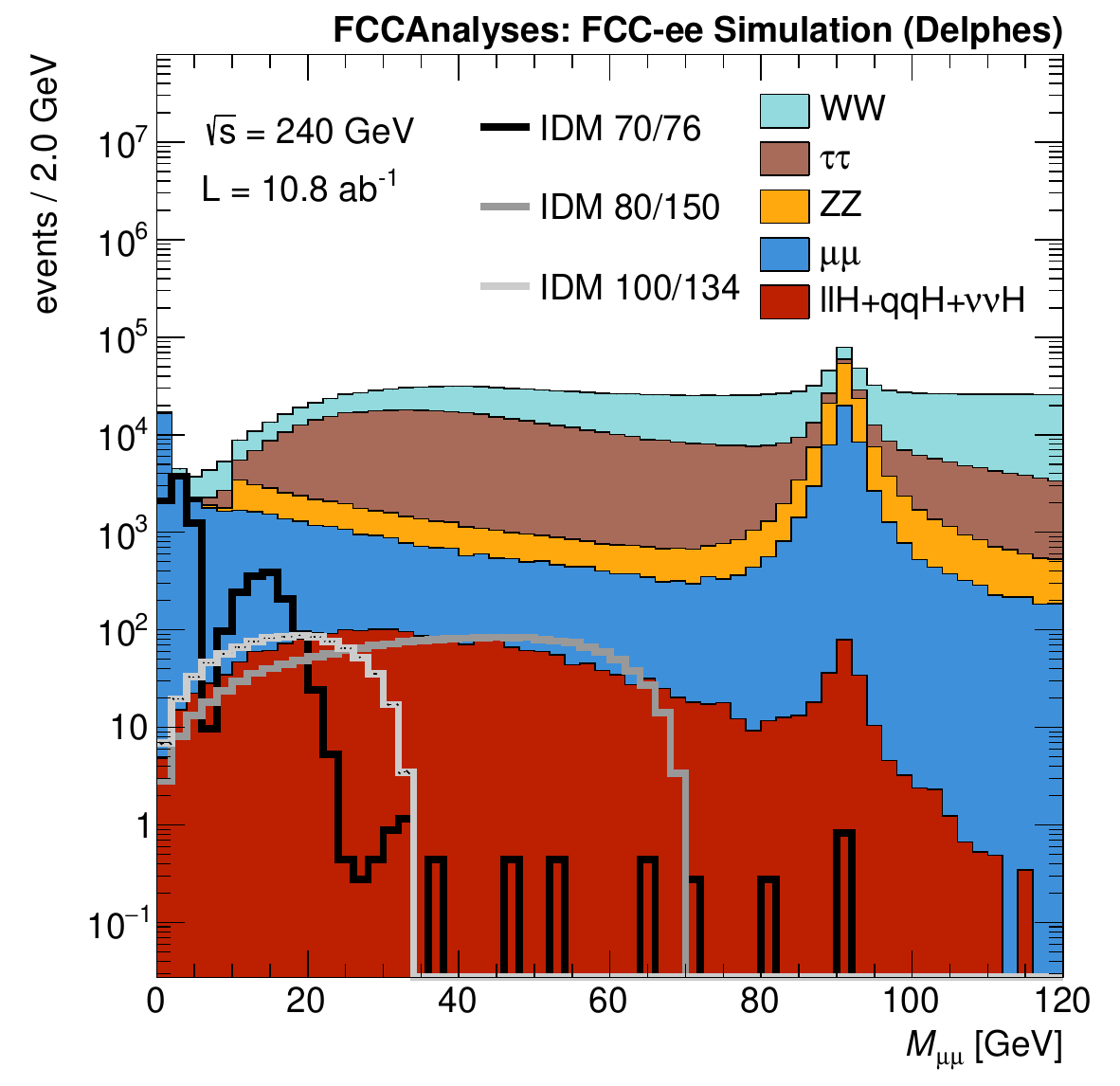}
\caption{Distribution of the dimuon pair energy $E_{\mu\mu}$ (left) and invariant mass $M_{\mu\mu}$ (right) for \sqs{240}, for the SM backgrounds and several selected signal mass points for the scenario S1. The signal mass points are labeled by IDM \mh/\ma\ with masses given in GeV.} 
    \label{fig:input1}
\end{figure*}

\begin{figure*}[h!]
\includegraphics[width=0.45\textwidth]{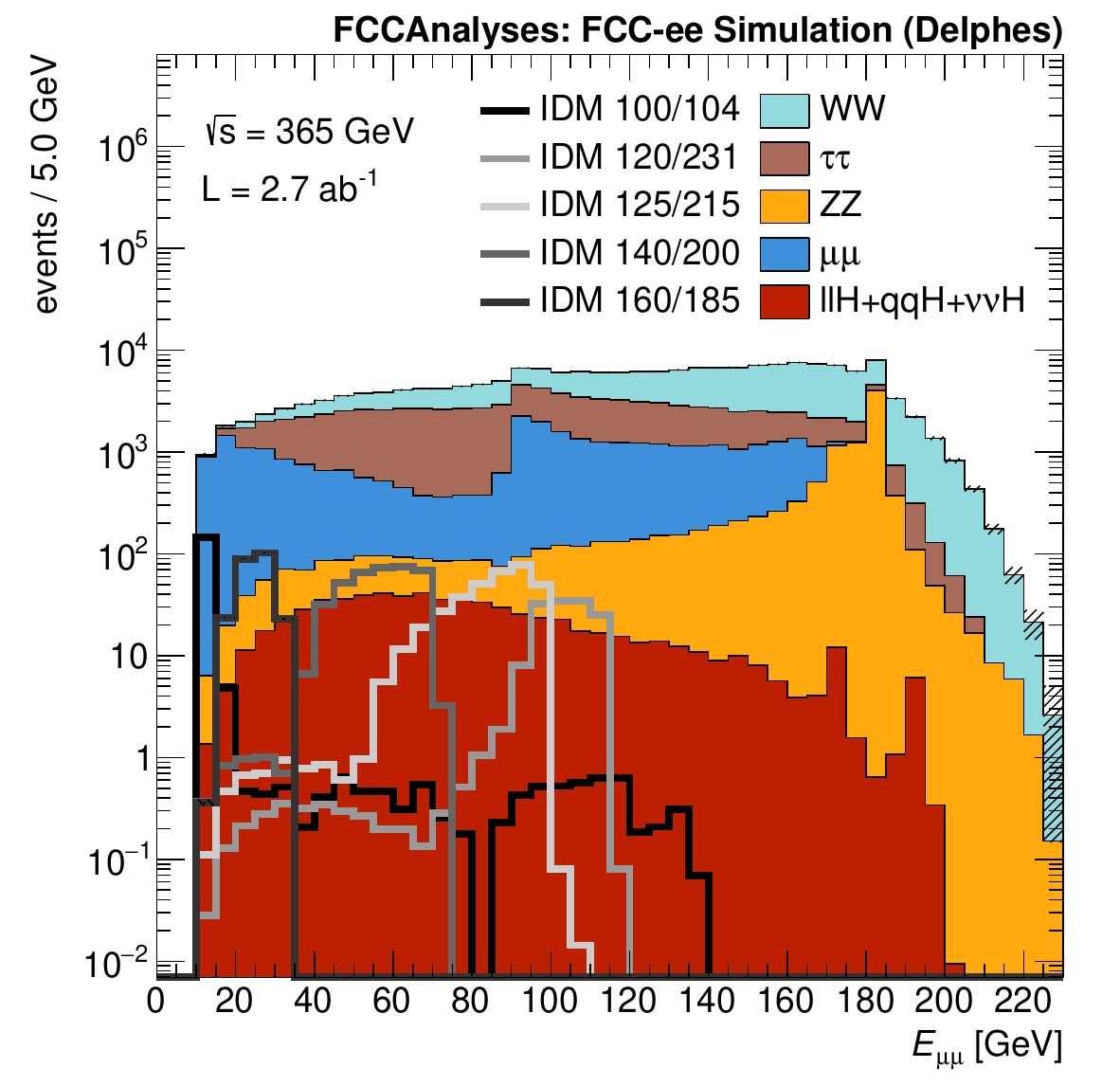}
\hfill
\includegraphics[width=0.45\textwidth]{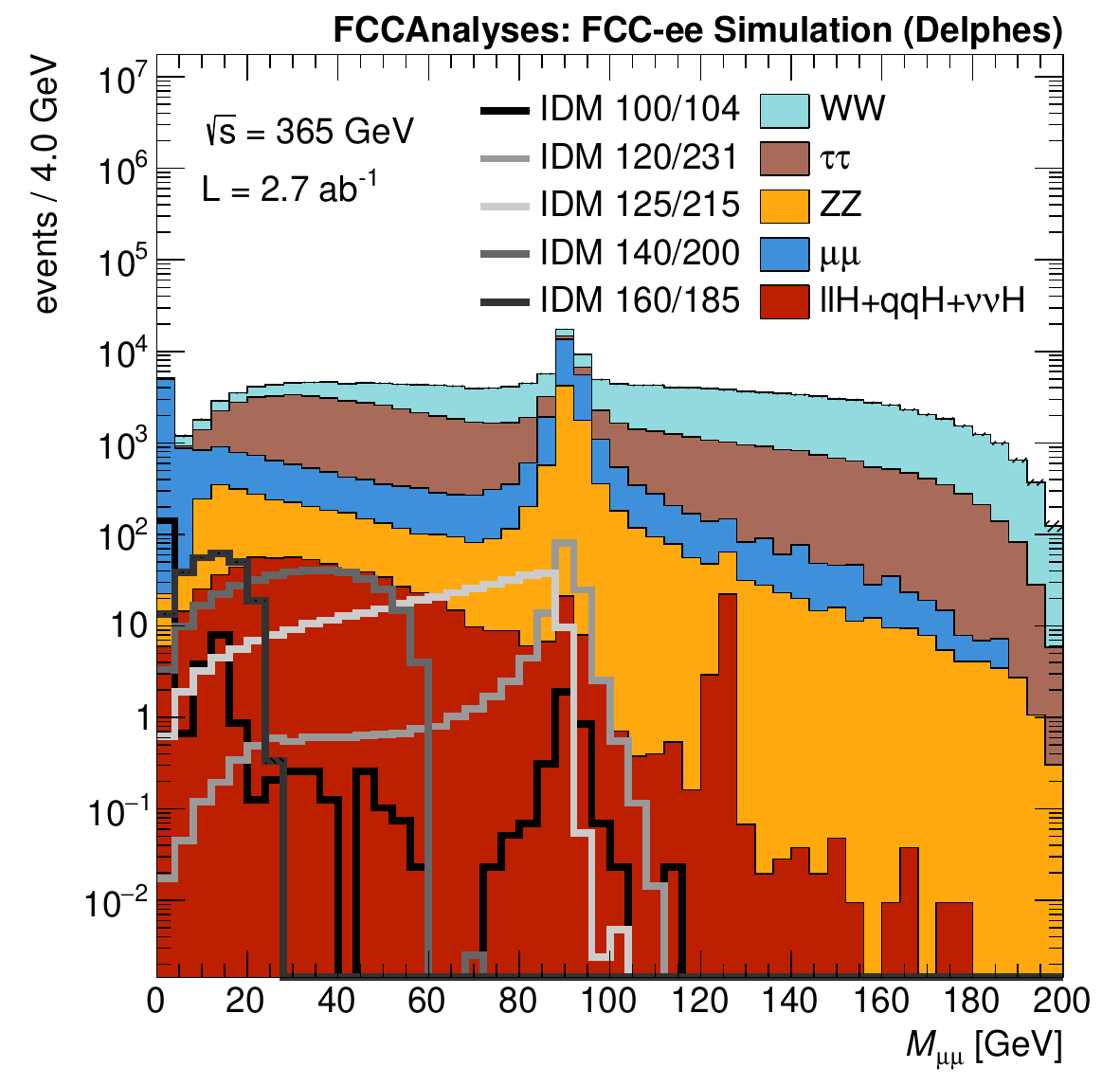}
\caption{Distribution of the dimuon pair energy $E_{\mu\mu}$ (left) and invariant mass $M_{\mu\mu}$ (right) for \sqs{365}, for the SM backgrounds and several selected signal mass points for the scenario S1. The signal mass points are labeled by IDM \mh/\ma\ with masses given in GeV.} 
    \label{fig:input2}
\end{figure*}

The dilepton pair $E_{\ell\ell}$ and invariant mass $M_{\ell\ell}$ distributions are shown in Figs.~\ref{fig:input1} and~\ref{fig:input2} for the different background processes and several representative signal points, for the $\mu\mu$ selection and at \sqs{240 and 365}, respectively. Note that the simulation curve shown for each signal point is the sum of both the $HH\ell\ell$ and $HH\ell\ell\nu\nu$ processes. In these distributions, large variations are seen in kinematic shapes between different signal points. Therefore, to take into account these differences, the network used is a parametric Neural Network~\cite{Baldi:2016fzo,Anzalone:2022hrt} (pNN) in which the two signal masses, \mh\ and \ma, are also used as input  along with the kinematic features listed above. By including the signal parameters as input, the model learns the optimal selection as a function of which mass hypothesis is being tested. As the signals can have different kinematic distributions, the trained pNN will apply a different set of selections that are optimal for each mass point. This allows for training of only a single network  per centre-of-mass energy whilst ensuring strong sensitivity to all signal points in the parameter space. For background events, which have no intrinsic IDM parameters associated with them, \mh\ and \ma\ are randomly chosen at the start of every training epoch from the set of IDM masses trained on. More details on pNNs and definitions of the relevant vocabulary can be found in Refs.~\cite{Baldi:2016fzo,Anzalone:2022hrt}.

The pNN is implemented in PyTorch~\cite{pytorch}, and contains four sequential feed-forward layers each with 250 units and a ReLU activation function, totaling roughly 200,000 learnable parameters. The last three layers are preceded by a dropout layer with a rate of $0.2$, and the network output is passed through a sigmoid function for binary classification. The loss function for the training of the model is a weighted binary cross-entropy loss function:
\begin{eqnarray}
   \lefteqn{ \text{Weighted BCE} = }  \\
   &&-\frac{1}{N} \sum_{i=0}^{N} w_i [ y_i \log(\text{pred}_i) + (1 - y_i) \log(1 - \text{pred}_i) ] \nonumber
\end{eqnarray}

where the sum is over the samples, $\text{pred}_i$ is the model prediction,
$y_i=\{0,1\}$ is the sample label, and $w_i$ is the sample weight. This is trained with the Adam optimiser with a learning rate of $0.0001$.

Each MC sample is split into three datasets: training, validation and test with fractions of the total dataset being 50\%, 20\% and 30\%, respectively. The training set is used to train the model, and the validation set is used to measure the performance of the model during training. Finally, the model's performance is evaluated using the test set.

Before training, events are re-weighted such that the sum of the signal weights equals the sum of the background weights. Moreover, the signals are re-weighted such that the sum of weights are equal for all IDM mass points to ensure that the training is not skewed towards certain signal points.

A single network is trained on both channels, $ee$ and $\mu\mu$, and all features that are passed to the pNN are standardised such that their mean is 0 and their standard deviation is 1. Training is conducted for 100 epochs, with early stopping if the validation loss does not decrease after 20 epochs. Finally, the model with the lowest validation loss is taken as the final model.  No overfitting between the training, validation and test sets is found.

The pNN can smoothly interpolate to mass points not trained on when applied to the background samples. The signal distributions at intermediate points are found using cubic spline interpolation. This involves a shape parameter $\varepsilon$ that is determined by minimising the leave-one-out cross-validation error~\cite{Rippa1999AnAF}. The interpolation capability of the setup was checked for several points of the parameter space, by comparing the results when those points are included or not in the training of the model. The pNN results were also compared with a simple implementation of a boosted decision tree using the same input features inside the TMVA framework~\cite{TMVA}, and comparable results were obtained on the individual IDM BPs.

\begin{figure*}[h!]
    \includegraphics[width=0.49\textwidth]{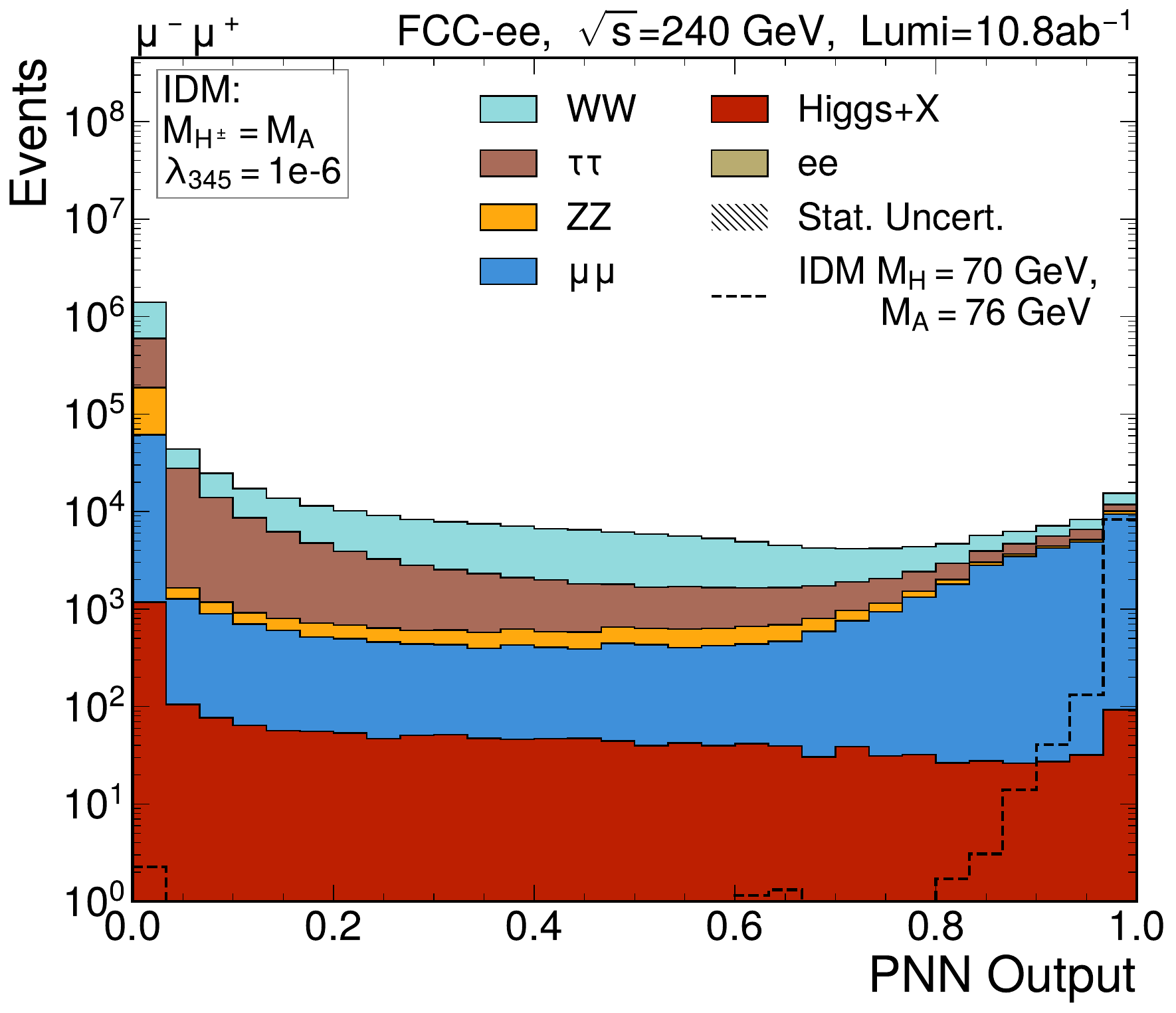}\hfill
    \includegraphics[width=0.49\textwidth]{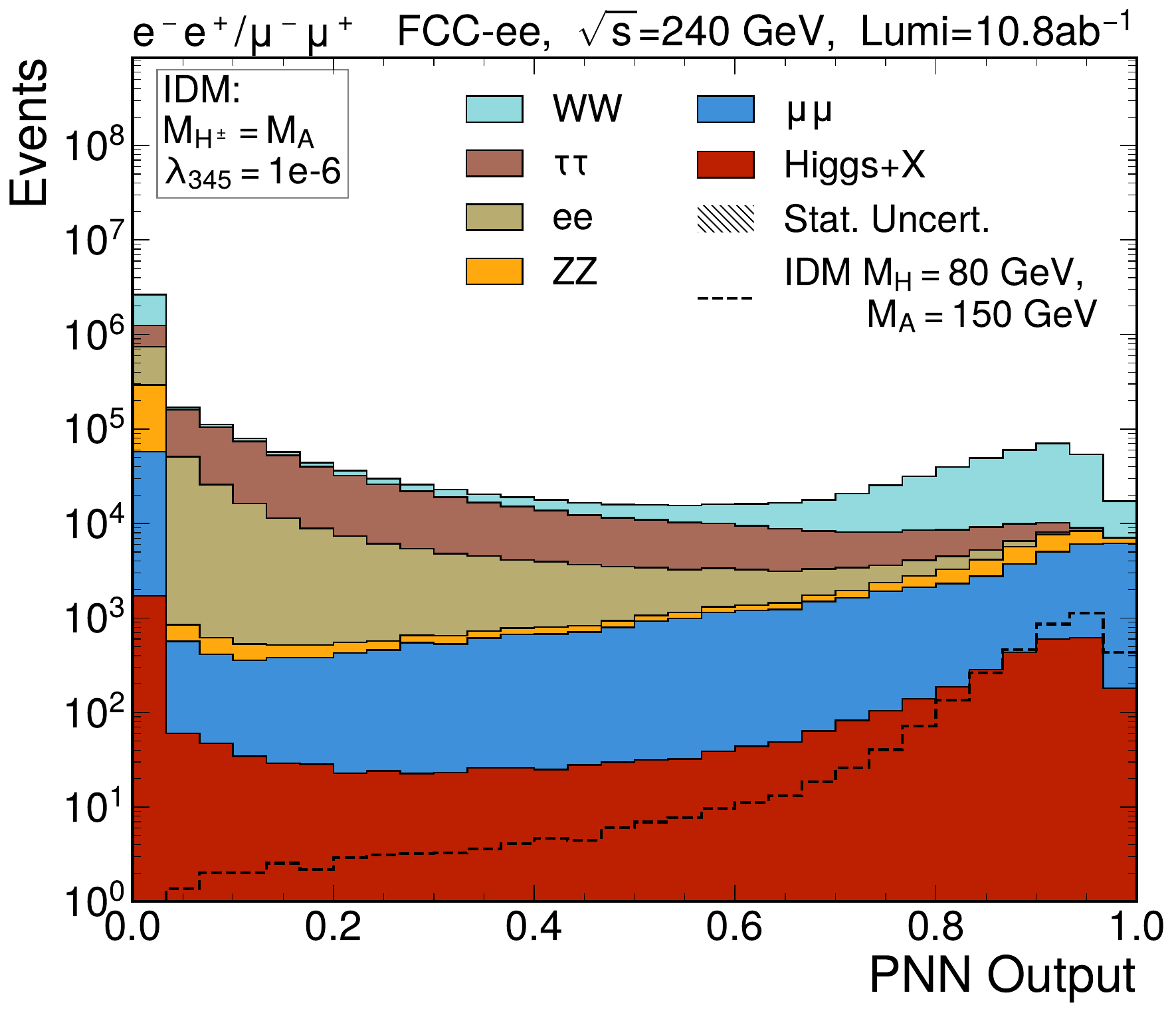}\\
    \includegraphics[width=0.49\textwidth]{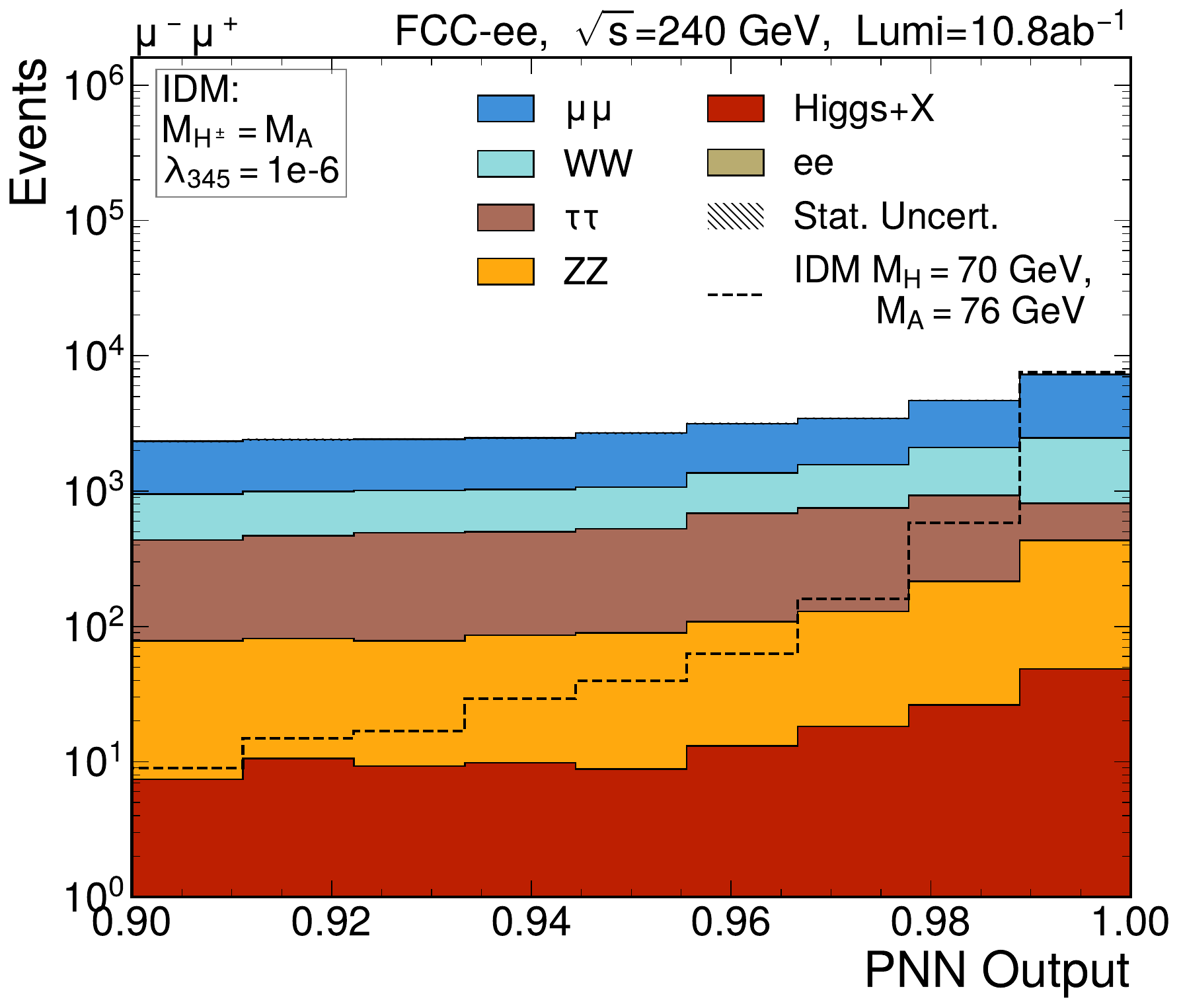}\hfill
    \includegraphics[width=0.49\textwidth]{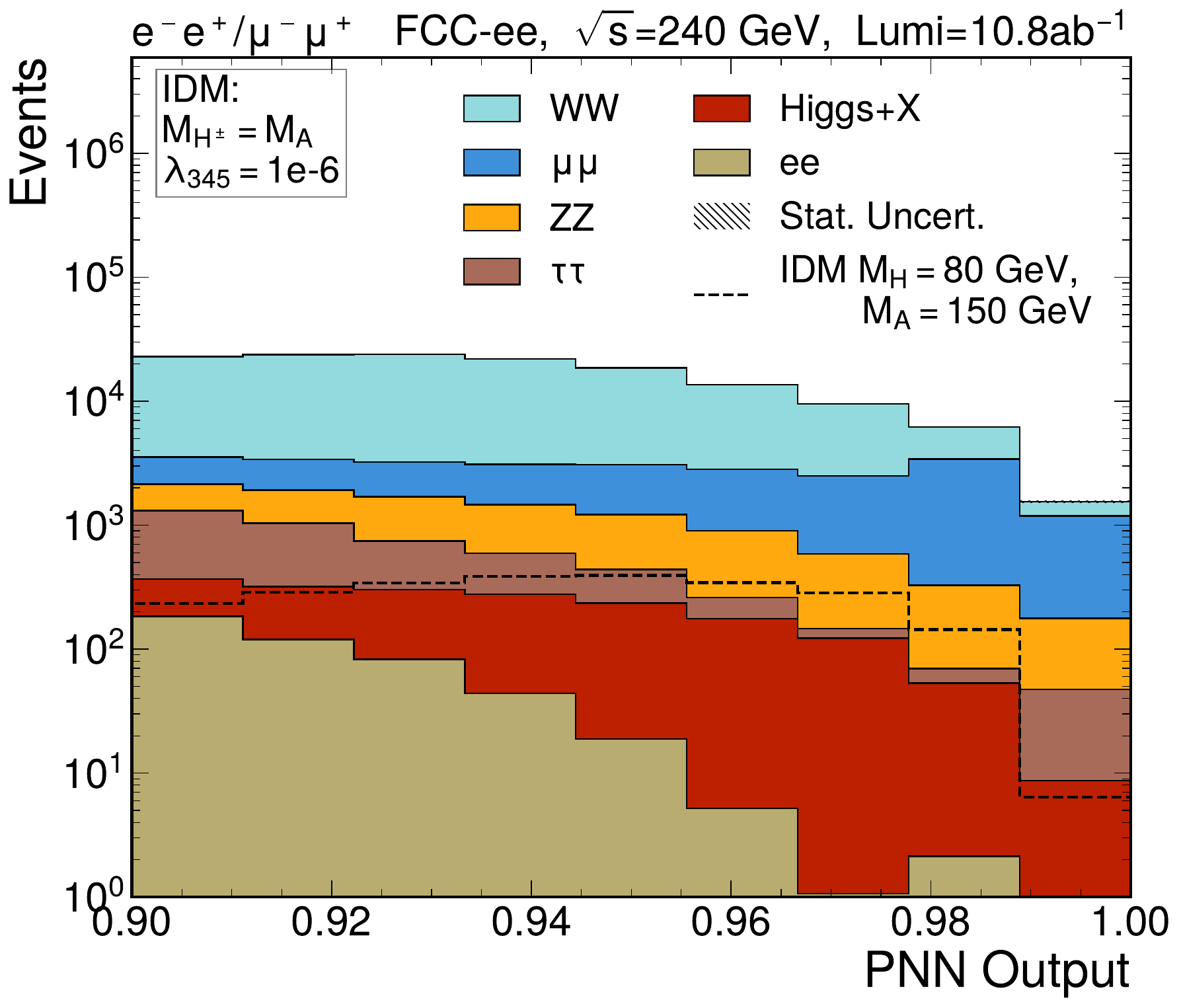}
    \caption{Output discriminator of the pNN for several signal points for \sqs{240} samples summing $ee$ and $\mu\mu$ contributions if \mdiff$>30$\,\GeV, only $\mu\mu$ elsewhere. Top: full-range distribution, bottom: selection defining the signal region for the extraction of the results. }
    \label{fig:pnn240}
\end{figure*}

\begin{figure*}[h!]
    \includegraphics[width=0.49\textwidth]{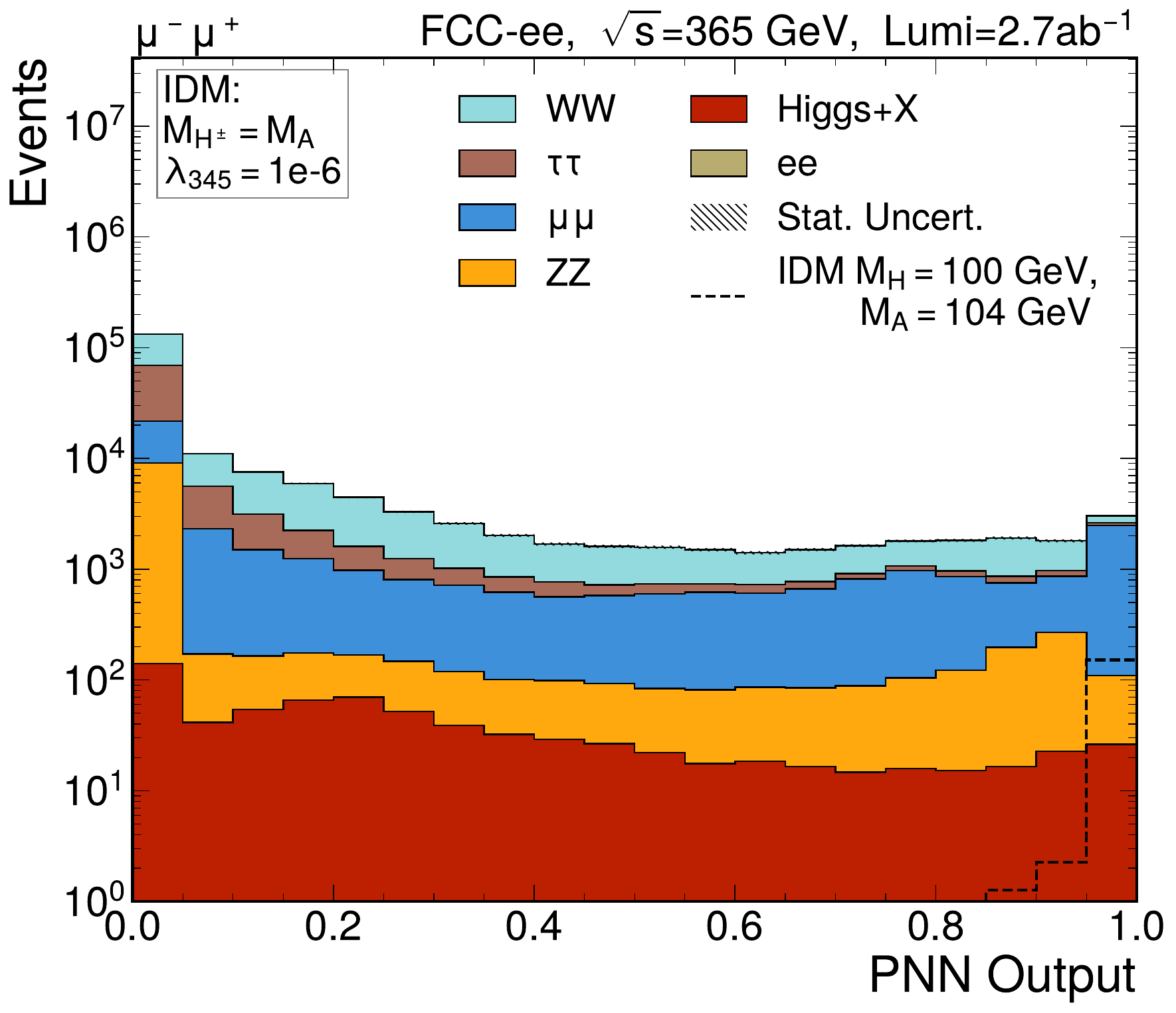}\hfill
    \includegraphics[width=0.49\textwidth]{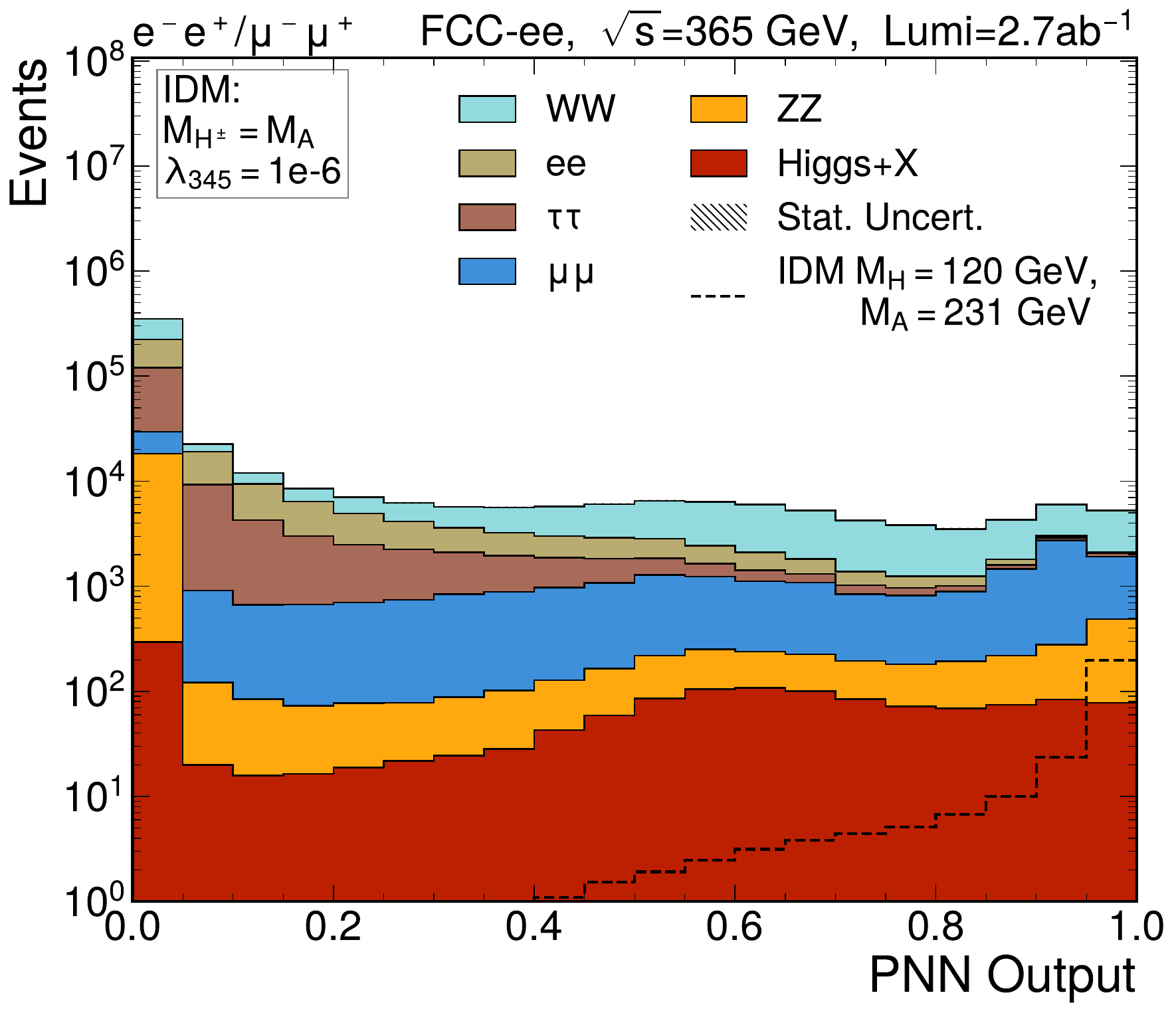}\\
    \includegraphics[width=0.49\textwidth]{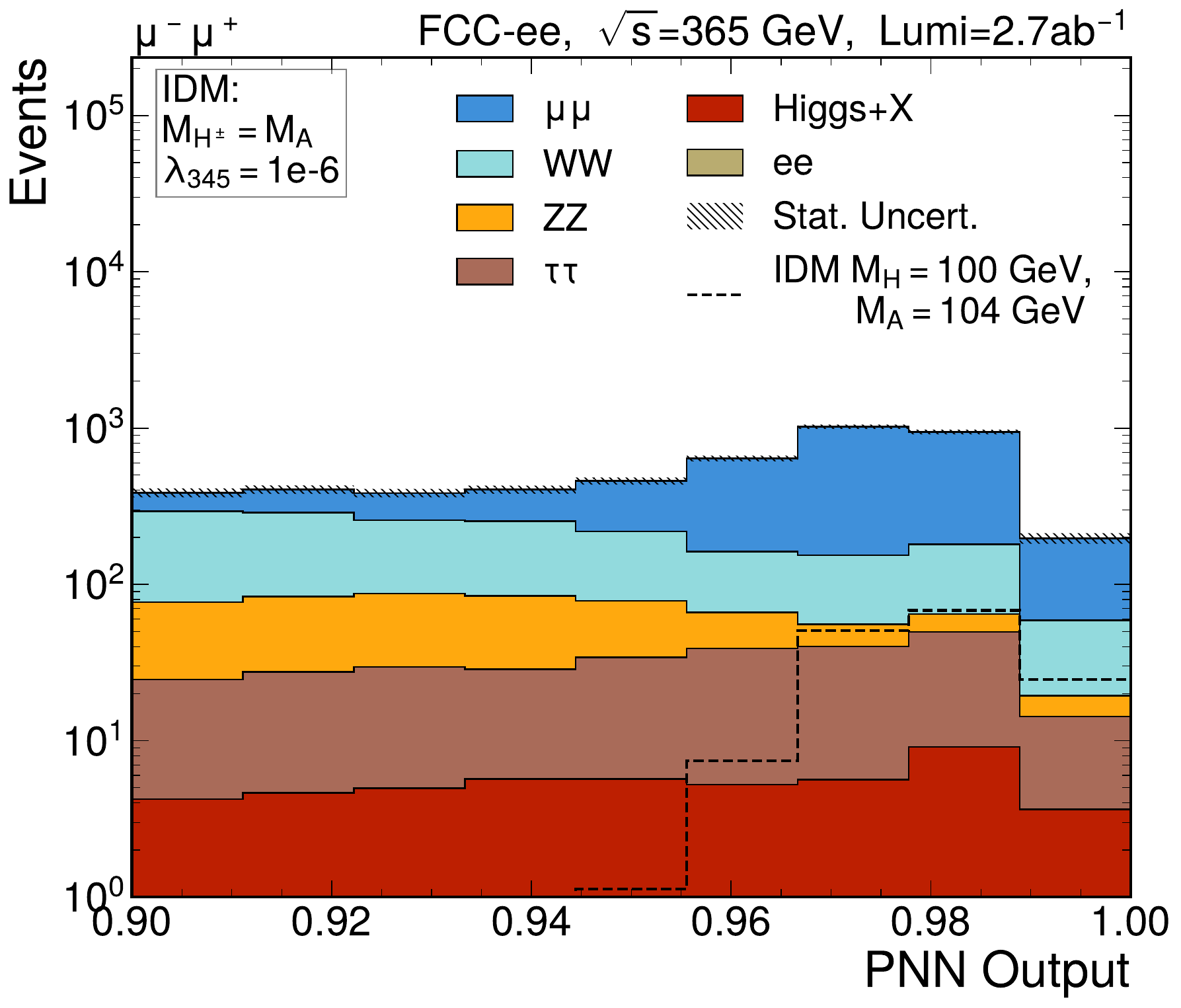}\hfill
    \includegraphics[width=0.49\textwidth]{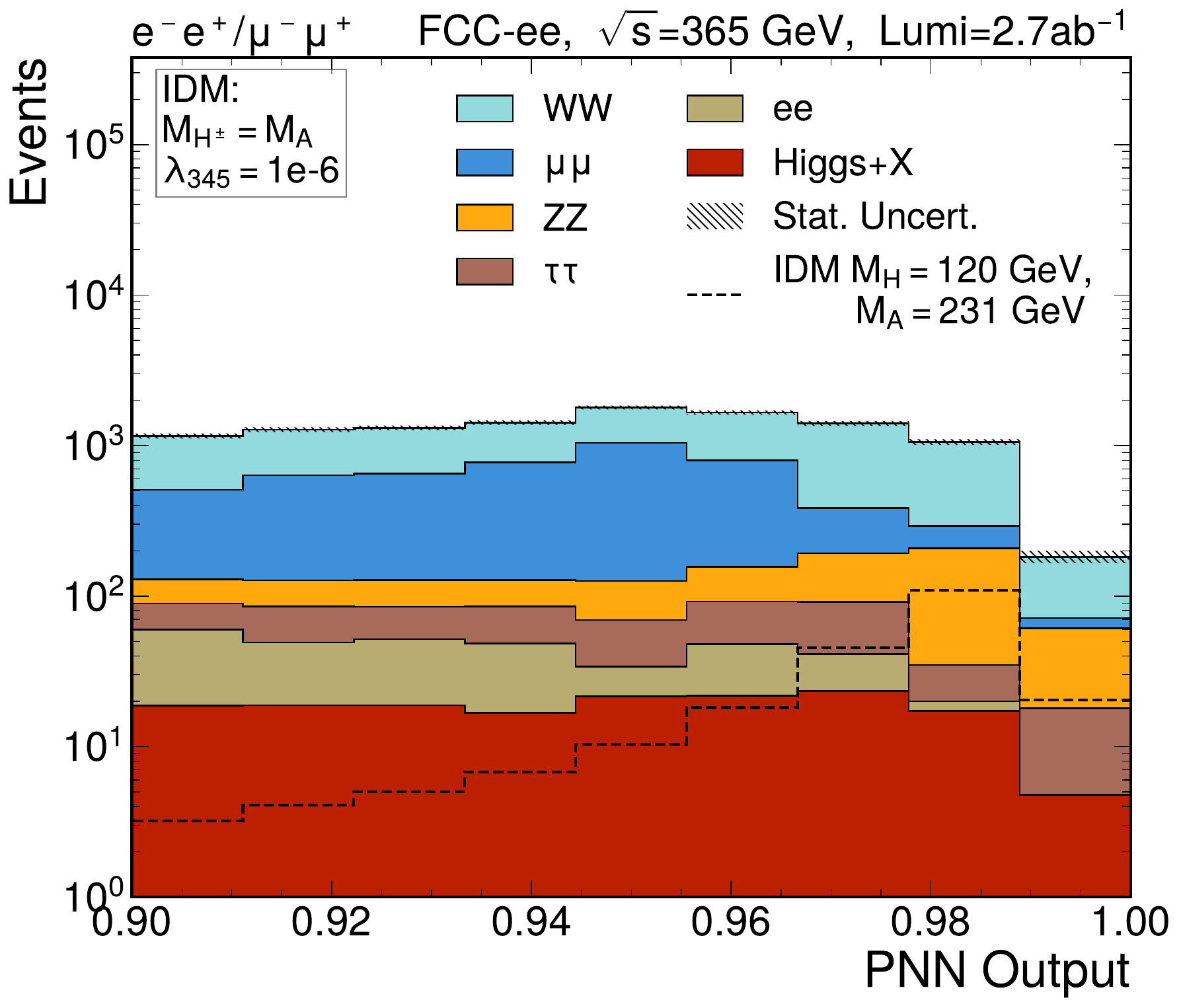}
    \caption{Output discriminator of the pNN for several signal points for \sqs{365} samples summing $ee$ and $\mu\mu$ contributions if \mdiff$>30$\,\GeV, only $\mu\mu$ elsewhere. Top: full-range distribution, bottom: selection defining the signal region for the extraction of the results.}
    \label{fig:pnn365}
\end{figure*}

The pNN distribution is shown for several representative signal mass points in Fig.~\ref{fig:pnn240} and~\ref{fig:pnn365} for \sqs{240 and 365}, respectively. The top row shows the full distribution whilst the bottom row shows the distribution above 0.9 which is used for extracting the results in the next section. The efficiency of the selection pNN $>$ 0.9 is above 50\% for all signal points, and above 90\% in the region $10\,<\, \mdiff\,<\,50\,\GeV$. The lowest efficiencies are obtained when \mdiff\ is close to the $Z$ boson mass, and at very low \mdiff.

\section{Results}
\label{sec:res}

A maximum likelihood fit of the pNN output above 0.9 is performed using the CMS Combine package~\cite{CMS:2024onh}.  The threshold of 0.9 is chosen to have maximum signal significance over backgrounds for all points. Both 95\% confidence level (CL) expected upper limits, assuming no signal for exclusion, and expected significance, assuming a signal rate predicted at a given parameter point for discovery, are calculated using profiled likelihood ratio test statistics, using asymptotic approximations for their distributions~\cite{Cowan:2010js}. A single source of uncertainties is included as nuisance parameters: the MC statistical uncertainties~\cite{Conway:2011in}. Future work should also include systematic uncertainties as additional nuisance parameters.

The $ee$ and $\mu\mu$ channels are fitted simultaneously. As the inclusive $ee$ production was simulated only for $M_{ee}>30$\,GeV, the two final states $ee$ and $\mu\mu$ are combined for \mdiff$>30$\,GeV, below which only the $\mu\mu$ result is shown. 

\begin{figure*}[h!]
    \includegraphics[width=0.45\textwidth]{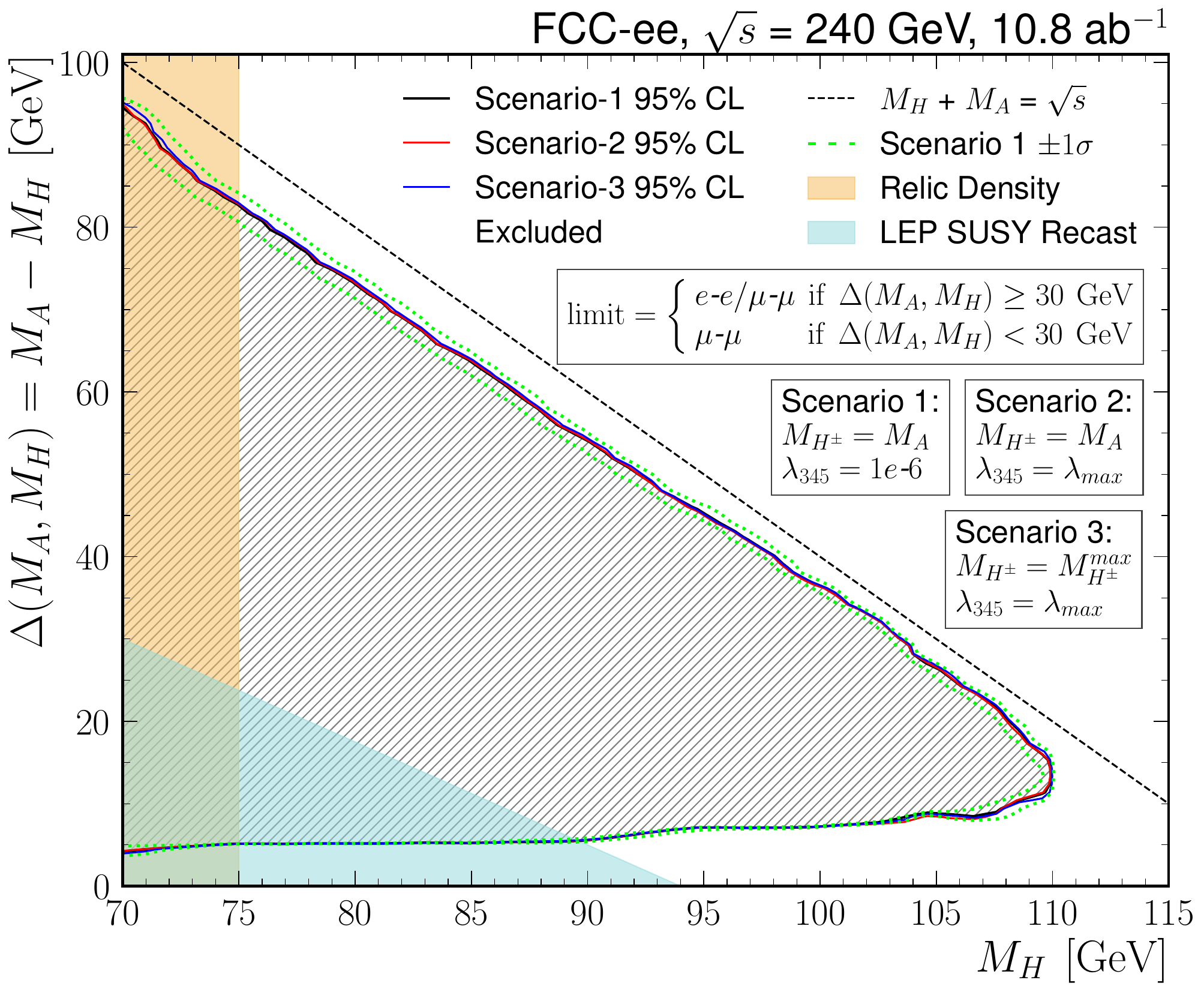} \hfill    
    \includegraphics[width=0.45\textwidth]{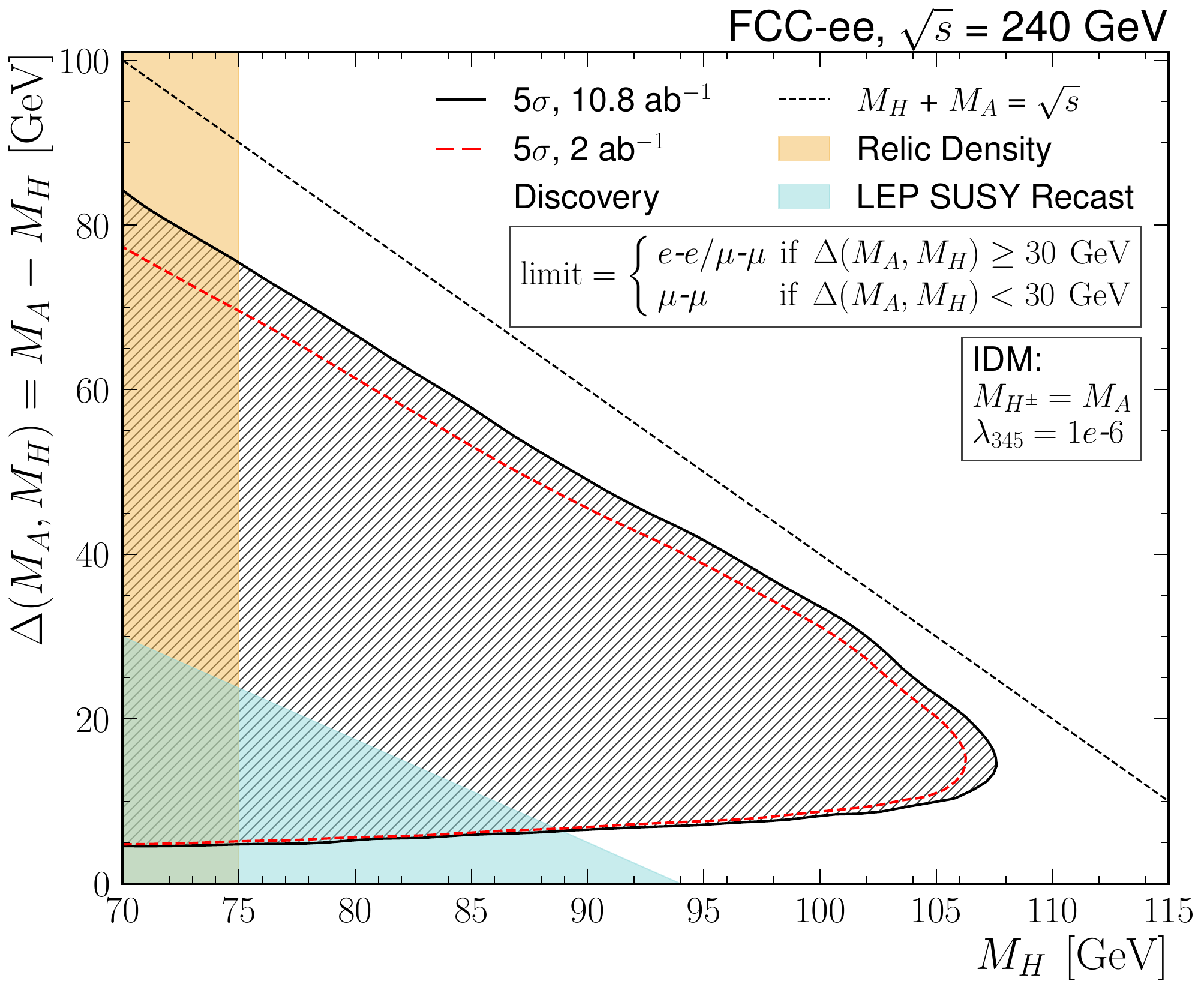}
    \caption{95\% CL exclusion contour (left) and $>5$-sigma discovery contour (right) in the \mdiff\ vs \mh\ plane, for \sqs{240} with total integrated luminosities of 10.8 and 2\ab.  Only statistical uncertainties have been included in the fit.}
    \label{fig:lim240}
\end{figure*}

 Contours are drawn for when the expected upper limit is smaller than the theoretical cross-section, giving the 95\% CL exclusion region shown in Fig.~\ref{fig:lim240} (left) in the \mdiff\ vs \mh\ plane, for \sqs{240}, for the total integrated luminosity of 10.8\ab. The three different scenarios defined in section~\ref{sec:idm} are shown as separate lines, well within the $\pm 1\sigma$ contour shown in green for scenario S1, including only the statistical uncertainties. The kinematic limit for on-shell production is highlighted by the dashed line. With the full luminosity of 10.8\ab, almost the entire parameter space can be excluded at 95\% CL, reaching \mh$ = 110$\,\GeV.

 Contours are also drawn for when the expected significance reaches $5$-sigma, giving the $>5$-sigma discovery region shown in Fig.~\ref{fig:lim240} (right) for \sqs{240}, in the scenario S1, for the total integrated luminosity of 10.8\ab\ and 2\ab. The regions excluded by relic density constraints and LEP SUSY recast (see section~\ref{sec:idm}) are overlayed\footnote{Note that in principle scenarios 2 and 3 are now excluded by direct detection constraints from \cite{LZ:2024zvo}. However, the main target here was to check the impact of switching on the respective channel with a 125 \GeV~ scalar mediator. We therefore chose to keep the corresponding results for illustration purposes.}. The discovery reach extends to  \mh$ = 108\,\GeV$ for \mdiff$=15$\,\GeV, for the nominal FCC luminosity scenario.  The other luminosity scenario of 2\ab\ is considered for comparison with a linear collider option at \sqs{250}~\cite{Altmann:2025feg}.

\begin{figure*}[h!]
    \includegraphics[width=0.45\textwidth]{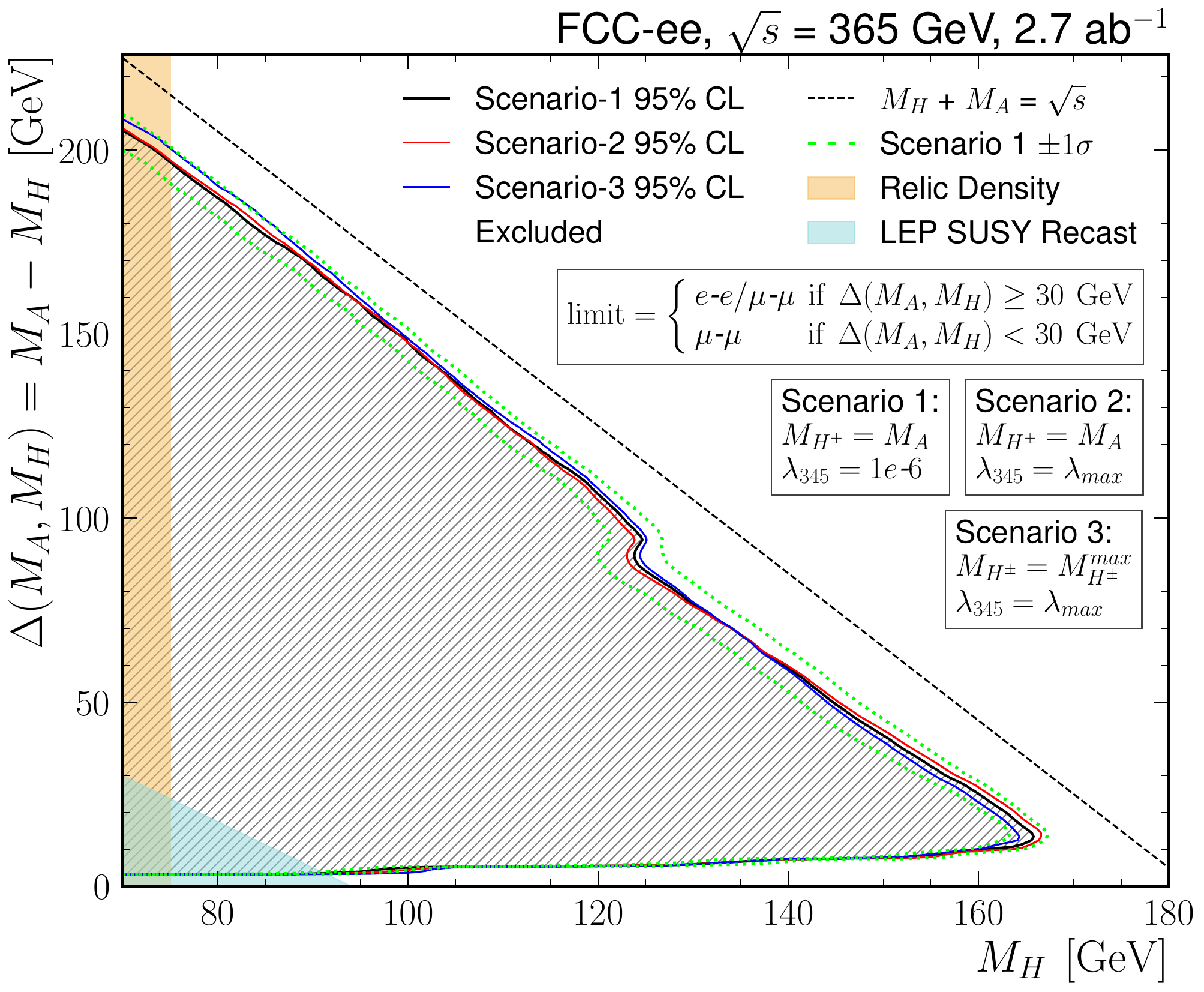} \hfill
    \includegraphics[width=0.45\textwidth]{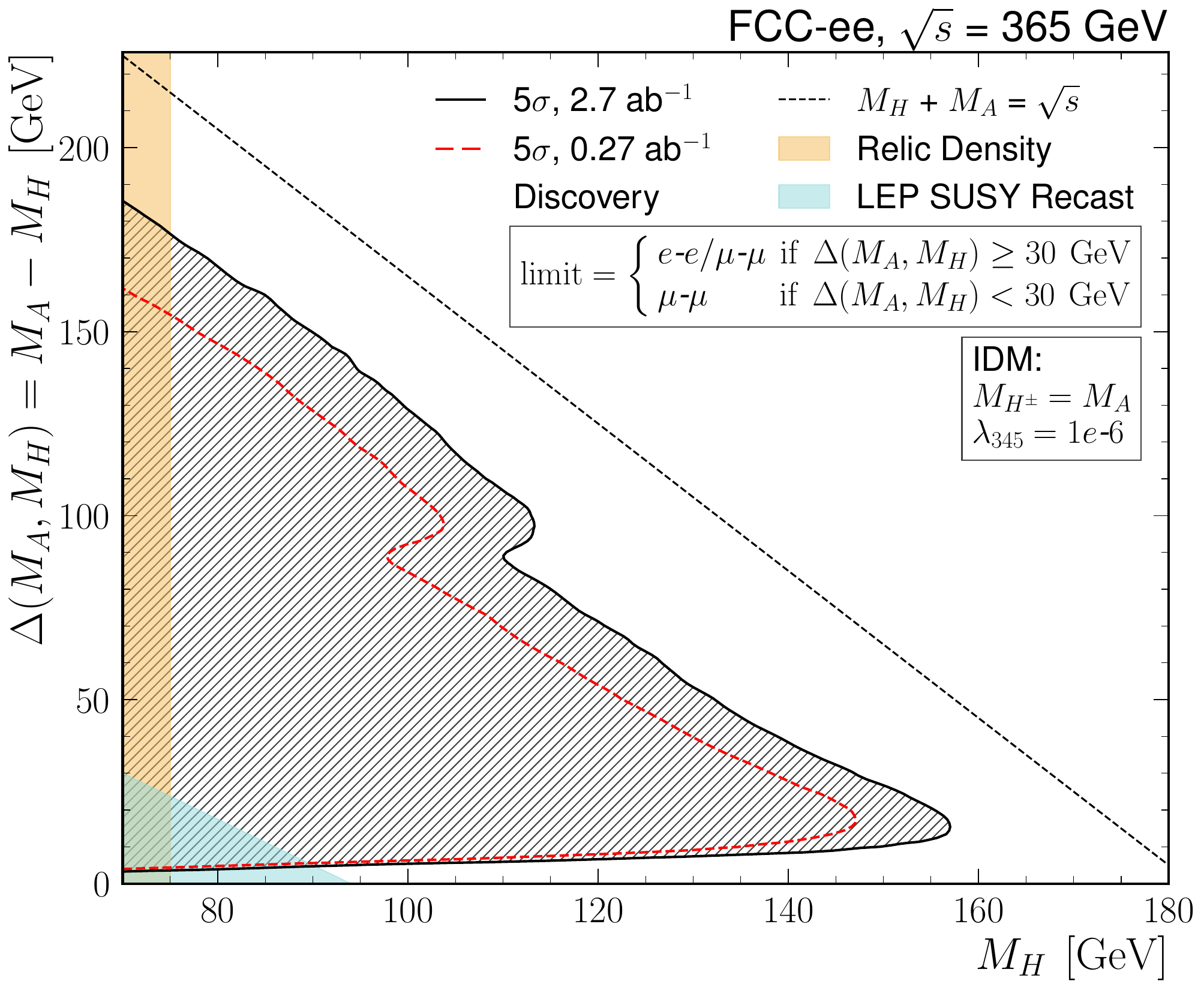}
    \caption{95\%CL exclusion contour (left)  and $>5$-sigma discovery contour (right) in the \mdiff\ vs \mh\ plane, for \sqs{365} with a total integrated luminosity of 2.7 ab$^{-1}$.  Only statistical uncertainties have been included in the fit.}
    \label{fig:signif365}
\end{figure*}

The 95\% CL exclusion (left) and $>5$-sigma discovery (right) regions are shown in Fig.~\ref{fig:signif365} for \sqs{365}, for the total integrated luminosity of 2.7\ab.  The regions excluded by relic density constraints and LEP SUSY recast are overlayed. As for \sqs{240}, the three different scenarios defined in section~\ref{sec:idm} are shown as separate lines and give results within the $\pm 1\sigma$ contour shown in green for scenario S1, including only the statistical uncertainties. The wedge in the distribution observed for a mass splitting around the on-shell Z boson mass is due to reduced sensitivity from reduced discrimination against background. The 95\% CL excluded region reaches \mh$=165$\,\GeV. The discovery reach extends to  \mh$ = 157$\,\GeV\ for \mdiff$=15$\,\GeV\ for the nominal FCC luminosity scenario, and is also shown in Fig.~\ref{fig:signif365} right for a scenario with ten times less luminosity , corresponding to a linear collider option at \sqs{350}~\cite{Altmann:2025feg}.

\section{Conclusion}
\label{sec:concl}

The Inert Doublet Model predicts additional scalars which couple only to bosons. The lighest neutral scalar, $H$, is also stable and provides an adequate dark matter candidate. 
The pair production of such new scalars is investigated in a final state containing two electrons or two muons, with $e^+e^-$ collisions at \sqs{240 and 365}, in the context of the future circular collider at CERN, FCC-ee.  A parametric neural network is trained to discriminate the different signal mass points against the backgrounds. A maximum likelihood fit of the output of the neural network is performed, including only statistical uncertainties. With a total integrated luminosity of 10.8\,(2.7)\,\ab\ for \sqs{240\,(365)}, the discovery reach is explored and covers almost the entire kinematic phase space available at \sqs{240\,(365)} reaching \mh$ = 108\,(157)\,\GeV$ for \mdiff$=15$\,\GeV. Furthermore, almost the entire parameter space available in the \mdiff\ vs. \mh\ plane is expected to be excluded at 95\% CL, reaching up to \mh$=110\,(165)$\,\GeV.  Parts of the 2-dimensional plane are already excluded by a recast of LEP results and constraints from relic density, i.e. overclosure of the universe. The FCC-ee reach will be complementary to that of higher-energy collisions from the LHC and HL-LHC, which probe an area with larger \mdiff. 

\backmatter

\bmhead{Acknowledgments}

TR has received support from grant number HRZZ-IP-2022-10-2520 from the Croatian Science
Foundation (HRZZ); in addition, this work was supported by the Science and Technology Facilities Council, UK. This research was supported in part by grant NSF PHY-2309135 to the Kavli Institute for Theoretical Physics (KITP). The authors thank A.F. Zarnecki for useful discussions. TR also wants to thank the Kavli Institute for Theoretical Physics (KITP) at the University of California, Santa Barbara, and in particular the organisers of the programme "What is particle theory ?", for their hospitality while parts of this work were completed.

\section*{Declarations}

\begin{itemize}
\item Competing interests: the authors have no relevant financial or non-financial interests to disclose.
\item Data availability: this manuscript has no data associated beyond references given in the text. 
\end{itemize}



\end{document}